\documentclass[%
 reprint,
 amsmath,amssymb,
 aps,
hidelinks]{revtex4-1}

\usepackage{xcolor}
\usepackage[normalem]{ulem}

\usepackage{graphicx}% Include figure files
\usepackage{dcolumn}% Align table columns on decimal point
\usepackage{bm}% bold math
\usepackage{texshade}
\usepackage{amsmath}
\usepackage{amssymb}
\usepackage{mathtools}
\usepackage{float}
\usepackage{soul}
\usepackage{tikz}
\usepackage{orcidlink}
\usetikzlibrary{shapes.geometric, arrows}

\begin{document}

%\preprint{APS/123-QED}

\title{General approach for partitioning and phase separation in macromolecular coexisting phases}% Force line breaks with \\

\author{Vikki Anand Varma\>\orcidlink{0000-0002-1646-1704}}
\affiliation{Department of Mechanical and Materials Engineering, University of Turku, Vesilinnantie 5, FI-20014 Turku, Finland}

\author{Alberto Scacchi\>\orcidlink{0000-0003-4606-5400}}
\email{alberto.scacchi@utu.fi}
\affiliation{Department of Mechanical and Materials Engineering, University of Turku, Vesilinnantie 5, FI-20014 Turku, Finland}

\date{\today}% It is always \today, today,

\begin{abstract} 

Partitioning of (bio)materials in polymeric mixtures is a key phenomenon both in cellular environments, as well as in industrial applications. In cells, several macromolecules are suspended within different biomolecular phases. On the other hand, the coexistence of polymeric aqueous phases has been exploited for the extraction and purification of (bio)materials suspended in water. Despite its relevance, key physical and chemical properties controlling the phase behavior of these complex systems are still lacking. Here, we have developed a classical density functional theory approach for describing the phase coexistence and partitioning of an arbitrary number of polymers and suspended materials. As a case example, we focus on a binary mixture of phase separating polymers in which a third material is dispersed. We explore the effect of size ratios and affinities between the different materials and address their distribution and coexisting densities, and find optimal conditions for partitioning. 

%Result shows that in mean-field approach integrated strength of potential serve the same purpose as the second virial coefficient in the experiment. Relation has been observed holding well at different polymer concentrations. For the same system with a fixed concentration, we observed that the partitioning of colloids increases linearly with the increase in the colloid-polymer affinity. Increasing polymers concentration increases the partitioning and get saturated at a maximum partitioning. It has also been observed that choosing the appropriate size of polymers rather than concentration can be the best strategy to get the optimum partitioning of colloids in the solution.  

\end{abstract}

\pacs{Valid PACS appear here}% PACS, the Physics and Astronomy
\maketitle

%\tableofcontents
\section{Introduction}

Partitioning within phase separating systems is fundamental to many everyday processes. Notable examples are cellular processes. Inside cells, liquid–liquid phase separation occurs alongside the presence of various biomaterials (colloidal or polymeric components), that are dispersed throughout the environment \cite{cell_seperation,doi:10.1021/jacs.3c10132, mejias2025liquid}. These components can partition differently depending on their chemical and physical properties, influencing cellular organization and function \cite{doi:10.1126/science.aaf4382}. Controlling partitioning within liquid-liquid phase separation also offers a platform for bioengineering \cite{D4PY90151G,doi:10.1021/acssynbio.3c00724}. Several developments for bio-inspired materials require the control of the partitioning of different species within liquid-phases \cite{https://doi.org/10.1002/adma.202414703}. On a larger scale, so-called aqueous two-phase systems have been proposed as an alternative to more conventional water purification methods  \cite{YAO2024111170,doi:10.1021/acs.langmuir.5c00749}. 

Despite the ubiquity of this process, a comprehensive understanding of the chemical and physical factors governing phase behavior in such complex systems remains limited, as responses often vary in a highly system-dependent manner. 

Coarse-grained particle-based simulations have recently been employed to investigate the partitioning of magnetic nanoparticles within a binary mixture of de-mixing polymers \cite{SCACCHI20251135}. This approach provides valuable insights into molecular-level interactions and self-organization within specific systems. However, when the goal is to systematically explore a broad landscape of chemically and physically distinct scenarios, such simulations become increasingly demanding in terms of computational time and resources. Because of this, we develop here an approach based on classical density functional theory (cDFT) \cite{Evans01041979, Evans01111993, evans1992density} able to describe the phase response of an arbitrary number of polymers and suspended materials. 

cDFT relies on the minimization of a free energy functional. A central challenge lies in identifying or constructing a functional that accurately captures the physical behavior of the system under study. Over the past several decades, significant effort has been devoted to both the development of new functionals and the refinement of existing ones (for reviews see \cite{Evans_2016, Vrugt02042020} and for recent developments, see e.g. \cite{PhysRevE.102.042140, 10.1063/5.0175065, PhysRevE.103.042103,Sammüller_2024, doi:10.1073/pnas.2312484120}). Here, we combine fundamental measure theory (FMT)~\cite{PhysRevLett.63.980}, free volume (FV) theory \cite{10.1063/1.1908765} and mean field (MF) theory \cite{10.1063/1.4993175} for obtaining the coexisting densities. In addition to using the standard MF (SMF) formalism as benchmark, we propose an enhanced MF (EMF) alternative that explicitly considers the effective free volume available to the different components. On the other hand, for obtaining inhomogeneous density profiles, we combine the Asakura-Osawa-Vrij (AOV) model \cite{PhysRevLett.85.1934} given for the model polymer within the framework of FMT, with our EMF formalism. We demonstrate the applicability of our approach by tackling a representative case from aqueous two-phase systems used in water purification from pollutants \cite{ASSIS2021121697}, as well as for biomaterial purification \cite{iqbal2016aqueous}, where two immiscible polymers phase-separate in the presence of a third dispersed material. Other relevant applications of these systems involve, e.g., the investigation of ultralow interfacial tension and micropattern formation in ferrofluids~\cite{rigoni2022ferrofluidic}, as well as their utilization in cell-mimicking microrobotic systems~\cite{furuki2024marangoni}. More broadly, our findings can be extended to other systems, such as biomolecular condensates~\cite{holehouse2025molecular} containing, e.g., dispersed globular proteins or virus-like particles.

Our work is structured as follows. In Sec. II A, we define the interaction potentials among different species. In Sec. II B, we introduce a general free energy functional for a multicomponent polymer-colloid mixture in bulk. In Sec. II C, we propose a general scheme to determine coexisting densities for a given multi-component and multiphase canonical ensemble. In Sec. II D, we extend the free energy functionals to address inhomogeneous systems. In Sec. III A we consider a specific case, namely two immiscible polymers and a dispersed colloid, and study its bulk response under different conditions, such as polymer size, affinity, and densities. Then, in Sec. III B, we study inhomogeneous properties, such as density distribution, of the specific case considered in bulk. In both Sec. III A and III B a brief comparison with particle-based simulations is presented.
Finally, in Sec. IV, we summarize our findings and discuss the limitations and possible extensions of our approach.

\section{Theory and implementation}

The present scheme relies on splitting any interaction potential into a soft (either attractive or repulsive) part and a hard-core repulsive part \cite{PhysRevA.5.939, 10.1063/1.1676556, henderson2021fundamentals, 10.1063/1.1674820}. The soft contribution is treated within a mean-field approximation, whereas the hard-core contribution can be handled by established methods within the cDFT framework, described in detail hereafter. 

\subsection{Interaction potential}

\subsubsection*{Polymer-polymer interaction}
Polymers are described by Gaussian potentials of the form\begin{equation}
\phi_{pq}(r)= \epsilon_{pq} e^{-\left(\frac{r}{R_{pq}}\right)^2},
\label{eq::pol-pol}
\end{equation}
where $p,q \in \{1, \dots, \rm n\}$, where n represent the total number of polymer species in the system. Throughout, all energies, including $\epsilon_{pq}>0$, are expressed in units of $k_{\rm B}T$, where $k_{\rm B}$ is the Boltzmann constant and $T$ the temperature of the system. $\epsilon_{pq}$ describes the repulsion strength between different types of polymers, controlling both species miscibility and polymer solvophilicities. Finally, $R_{pp}$ is a measure of the radius of gyration of each polymer species. To calculate the inter-species radii, we use the standard quadratic mean relation
\begin{equation}
    R_{pq}^2 = \frac{R_{pp}^2+R_{qq}^2}{2} ,
\label{equ.meanpolymer_size}
\end{equation}
which is based on the convolution of Gaussian functions.

%The radius of gyration of polymer strand is defined as,

%\begin{equation}
%R_g^2 = \frac{1}{N} \sum_{i=1}^N \left\langle \left( \mathbf{r}_i - \mathbf{R}_{\text{cm}} \right)^2 \right\rangle ,
%\label{eq.radius_of_gyration}
%\end{equation}
%where the summation is performed over the constituent particle of %the polymer strand and $R_{\rm cm}$ is the center of mass of the %polymer. 

%All the length is expressed in the unit of $\sigma$. 
The potential in Eq.~(\ref{eq::pol-pol}) is well suited to describe the effective interaction between both linear polymers and dendrimers in good solvents~\cite{Louis20002522,Bolhuis20014296, LIKOS2001267, Götze20047761,10.1063/1.2172596}. This and similar interaction potentials have been used extensively in the context of cDFT for polymers (see e.g. \cite{PhysRevE.64.041501, PhysRevResearch.3.L022008, 10.1063/5.0053365}).

\subsubsection*{Colloid-colloid interaction}
Colloidal particles are modeled as hard spheres, i.e. 
\begin{equation}
\phi_{cd}(r)=
\begin{cases}
+ \infty \hspace{0.5cm} &r < \sigma_{cd} \\
0 \hspace{0.5cm}  &r \geq \sigma_{cd},
\end{cases}
\label{eq:coll-coll}
\end{equation}
where $c,d \in \{\rm n+1, \dots, \rm n+m \}$, with $\rm m$ representing the total number of colloidal species. $\sigma_{cc}$ is the diameter of the spherical colloids, while the cross-diameter $\sigma_{cd}=(\sigma_{cc}+\sigma_{dd})/2$ (additive colloidal mixture). 

An attractive interaction between colloids can, in principle, be included within a mean-field treatment. For simplicity, however, we omit it here.

\subsubsection*{Colloid-polymer interaction}

The colloid-polymer interaction is split into two parts. The first models depletion forces via a hardcore repulsive potential, reading 
\begin{equation}
\phi_{cp}^{\rm rep}(r)=
\begin{cases}
+ \infty,  &r < \sigma_{cp}, \\
0,  &r \geq \sigma_{cp},
\end{cases}
\label{eq:coll-pol_rep}
\end{equation}
where $\sigma_{cp}=(\sigma_{pp}+\sigma_{cc})/2$, where $\sigma_{pp}$ represents the excluded volume effect of the polymers in the presence of colloids (see discussion on {\it free energy for hardcore interaction} in Sec. II B). The second, describing surface attraction (e.g. due to electrostatic or solvophobic interactions), is defined as 
\begin{equation}
\phi_{cp}^{\mathrm{att}}(r)=
\begin{cases}
0,                     &r  < \sigma_{cp},\\
-\epsilon_{cp},  &\sigma_{cp} \le  r  < \delta\sigma_{cp},\\
0,                     &r  \ge \delta\sigma_{cp}.
\end{cases}
\label{eq:coll-poll_att}
\end{equation}
where $\delta>1$ sets the range and $\epsilon_{cp}$ the strength of the interaction, respectively.

\subsection{Free energy functional for bulk system}
We start by defining the Helmholtz free energy functional per unit volume of a system with n polymers and m colloids. Generally, it can be written as the sum of the ideal and the excess contributions~\cite{Evans01041979}, i.e.
\begin{equation}
\mathcal{F}[\{\rho_i\}] = \mathcal{F}_{\rm id}[\{\rho_i\}]+\mathcal{F}_{\rm exc}[\{\rho_i\}], 
\label{eq::total_free_energy}
\end{equation}
where the ideal term takes the known form 
\begin{equation}
\beta\mathcal{F}_{\rm id}[\{\rho_i\}]=\sum _{i} \left[\rho_i \ln\left(\Lambda_i^3\rho_i\right) - \rho_i \right], 
\label{eq::ideal_free_energy}
\end{equation}
where $i \in \{1,..., \rm n+m\}$, $\rho_i$ is the number density of species $i$, $\beta=(k_{\rm B}T)^{-1}$, and $\Lambda_i$ is the de Broglie wavelength.  We express the excess part of the free energy as
\begin{equation}
\beta \mathcal{F}_{\rm exc}[\{\rho_i\}] = \mathcal{F}_{\rm HC}[\{\rho_i\}] + \mathcal{F}_{\rm FV}[\{\rho_i\}] + \mathcal{F}_{\rm EMF}[\{\rho_i\}]. 
\label{eq.4_8}
\end{equation}
The first term, $\mathcal{F}_{\rm HC}$, describes all hardcore contributions which rise from colloid-colloid interactions, Eq. (\ref{eq:coll-coll}). In this case, the only relevant densities correspond to $i\in \{\rm n+1, \dots, n+m \}$. The free volume term, $\mathcal{F}_{\rm FV}$, takes care of the depletion forces arising from colloid-polymer interactions (Eq.~(\ref{eq:coll-pol_rep})). Finally, the enhanced mean field term, $\mathcal{F}_{\rm EMF}$, describes the interactions between all polymers (Eq.~(\ref{eq::pol-pol})), as well as of the attractive contribution of the colloid-polymer interactions, Eq.~(\ref{eq:coll-poll_att}). %However, the perturbative approaches have serious implications due to the effective density of the polymers increasing due to the presence of colloidal particles. There are various ways to compensate for the loss in pressure and energy in the perturbative mean-field approach, which we will discuss in Sec. II (C). 

\subsubsection*{Free energy for hardcore interaction}

For calculating the coexisting densities of a phase separating system, one can focus on expressing the free energy for homogeneous systems. In this case, $\mathcal{F}_{\rm HC}$ can be expressed using the Rosenfeld's approximation~\cite{PhysRevLett.63.980}, which is very accurate within low density regimes. The latter reads \cite{Roth_2010}
\begin{equation}
\begin{aligned}
   \beta\mathcal{F}_{\rm HC}[\{\rho_i\}] &=\sum_{i=\rm n+1}^{\rm n+m} \left[- n_0^i \ln{\left(1-\sum_{j = \rm n+1}^{\rm \rm n+m} n_3^j\right)}\right] \\&+
    \sum_{\substack{i={\rm n+1}\\j={\rm n+1}}}^{\rm m+n} \frac{n_1^in_2^j}{1-\sum_{k=\rm n+1}^{\rm n+m} n_3^k} \\
   &+ \frac{1}{24 \pi} \sum_{i,j,k={\rm n+1}}^{\rm m+n }\frac{n_2^{i} n_2^{j} n_2^{k}}{\left(1- \sum_{l={\rm n+1}}^{\rm n+m} n_3^l\right)^2},
\end{aligned}
    \label{eq.rosenfeld}
\end{equation}
where $n_0, n_1, n_2$ and $n_3$ are weighted scalar densities, which for a homogeneous system is given by

\begin{equation}
    \begin{aligned}
        n_0^c &=\rho_c,\\
        n_1^c &=R_c \rho_c,\\
        n_2^c &= 4\pi R_c^2 \rho_c,\\
        n_3^c &= \frac{4}{3} \pi R_c^3 \rho_c.
    \end{aligned}
\end{equation}
Polymers act as depletants for all colloidal species due to an effective reduction of the accessible volume. This effect can be included by means of the FV theory \cite{10.1063/1.1908765}. The latter considers the cost of inserting a polymer at a given chemical potential for a specific density of colloids, and is given by
\begin{equation}
 \beta  \mathcal{F}_{\rm FV}[\{\rho_i\}] = - \sum_p \rho_p  \log{\alpha_p},
    \label{eq.freevolume}
\end{equation}
where $\alpha_{p}$ depends upon the polymers' and colloids' shape and size. For a particular total colloidal packing fraction $\eta = \sum_c\eta_c = \sum_c\rho_{\rm c}\sigma_{cc}^3\pi/6$, the general expression for $\alpha_p$ is given by \cite{10.1063/1.1908765}
\begin{equation}
\begin{aligned}
\alpha_p = (1-\eta) \exp\Bigg[ 
& -  \frac{A_p(\{\rho_c\})}{(1-\eta)}   -  \frac{B_p(\{\rho_c\})}{(1-\eta)^2}  \\
& - \frac{C_p(\{\rho_c\})}{(1-\eta)^3} 
\Bigg].
\end{aligned}
\label{eq4_12}
\end{equation}
The coefficients $A_p(\{\rho_c\})$, $B_p(\{\rho_c\})$ and $C_p(\{\rho_c\})$ for the polymer species $p \in \{1,\ldots, \rm n\}$ are given by
\begin{equation}
A_p(\{\rho_c\})= s_p\sum_ca_c \rho_c   + a_p \sum_c s_c \rho_c + v_p\sum_c \rho_c,
\label{eq.4_13}
\end{equation} 
\begin{equation}
\begin{split}
B_p(\{\rho_c\})&= \frac{1}{2} s_p^2 \left(\sum_c\rho_ca_c\right)^2 \\&+ v_p \left( \sum_c s_c \rho_c \right)   \left( \sum_c a_c \rho_c \right),
\end{split}
\label{eq.4_14}
\end{equation} 
and
\begin{equation}
C_p(\{\rho_c\})= \frac{1}{12\pi}v_p (\sum_c a_c \rho_c)^3,
\label{eq.4_15}
\end{equation} 
where $a_i = \pi \sigma_{ii}^2$ represents the surface area, $s_i = \sigma_{ii}/2$ the radius and $ v_i = \frac{\pi}{6} \sigma_{ii}^3$ the volume of the particle, with $c\in \{\rm n+1,\ldots, {\rm n+m} \}$ and $i\in \{1,\ldots,{\rm n+m}\}$. $\sigma_{cp}$ corresponds to the minimum distance between the center of mass of polymer $p$ and the colloid $c$, which represents the excluded volume between a colloid and a polymer. Note that $\sigma_{pp}$ is often chosen as $\sigma_{pp} = R_{pp} + \frac{1}{2} \sigma_{cc}$ \cite{PhysRevLett.85.1934}.

%However, a more precise assumption is provided by the Barker–Henderson (BH) prescription~\cite{10.1063/1.1701689}. For $\phi_{cp}(r)$ defined by Eq.~(\ref{eq::coll-pol_full}), the diameter corresponding to the excluded volume can be mapped to
%\begin{equation}
%\sigma_{\mathrm{BH}} = \int_0^{\sigma \lambda^{1/(\rm v-u)}} \Big[ 1 - e^{-\beta \phi_{cp}(r)} \Big] dr.
%\label{eq::bh_radius}
%\end{equation}
%Assuming $\sigma_{pp} = \sigma_{\rm BH}$ provides an accurate representation of the effective excluded volume while keeping the system description consistent with the underlying interaction potential~\cite{10.1063/5.0149865}.

\subsubsection*{Free energy for soft interaction}
In the presence of colloids, polymers experience confinement due to the excluded volume arising from the presence of colloids. This is particularly important when the total amount of colloids is at relatively high value. As a consequence, the perturbative mean field can underestimate the pressure and the free energy. Therefore, the effective polymer density has to be modified by a factor $\alpha'_p$, with $\alpha'_p \neq \alpha_p$ \cite{brader2003statistical, schmidt2003fluid, reiss1992statistical, warren1995effect}. Specifically, $\alpha_p$ is the accessible volume, which accounts for the restriction of the polymer center of mass within the excluded volume around colloids, while $\alpha'_p$ represents the actual volume available for the polymer body to occupy. Unlike $\alpha_p$, the form of $\alpha_p'$ is not known. %Especially in the case of colloid polymer interaction, most of the available literature rely on the void estimation in the limit of zero polymer density in a sea of colloids. However in the presence of affinity between colloid and polymer, that void estimation can lead to serious problem as now excluded volume increases compared to purely repulsive interactions.   
%
%In the presence of colloids, polymers experience confinement arising from the excluded volume generated by colloidal particles. This effect becomes particularly significant at relatively high colloid concentrations. As a consequence, the effective polymer density must be corrected by a factor $\alpha'_p$. Here, we choose $\alpha'_p \neq \alpha_p$. As $\alpha_p$ denotes the accessible volume fraction that accounts for the restriction of the polymer center of mass due to colloidal exclusion, whereas $\alpha'_p$ represents the actual volume fraction available for the polymer body itself. Unlike $\alpha_p$, the explicit form of $\alpha'_p$ is not known.  
%
%In the context of colloid–polymer interactions, most existing studies estimate the available void volume in the limit of vanishing polymer density within a colloidal suspension. However, in systems with finite affinity between colloids and polymers,
%
The simplest estimate of $\alpha'_p$ can be obtained by writing the effective volume in terms of an expansion over $\eta$, i.e.
\begin{equation}
V_p^{\rm eff} = V (1- \beta_1 \eta+\beta_2 \eta^2 +...) ,
\label{eq.veffective}
\end{equation}  
where $\eta$ is the total packing of the all colloids. In analogy with a van der Waals-type approximation~\cite{MELNYK2022120672}, we write the effective free volume available to polymers as  
\begin{equation}
V_p^{\rm eff} = V - \sum_c \left[\tfrac{1}{2} v_{cp} N_{c} - \tfrac{3}{8} v_{cp}^2 N_c^2 \right],
\label{eq.veff}
\end{equation}  
where $v_{cp}$ is the apparent volume arising from steric interactions between colloid $c$ and polymer $p$, given by $v_{cp} = \tfrac{\pi}{6} \sigma_{cp}^3$. Equation (\ref{eq.veff}) is obtained by considering $\beta_1 = \frac{1}{2}$ and $\beta_2 = \frac{3}{8}$ in Eq.~(\ref{eq.veffective}), while replacing $v_{cc}$ (representing the volume of colloid $c$ in $\eta_c$) with $v_{cp}$. The latter assumption allows us to also take into account the size of the polymers explicitly. Under this assumption, the corresponding effective polymer density is  
\begin{equation}
\rho'_p = \frac{N_p}{V_{\rm eff}} 
= \frac{N_p}{V- \sum_{c} \left(\tfrac{1}{2} v_{cp}N_c -\tfrac{3}{8}v_{cp}^2 N_c^2 \right) } 
= \frac{\rho_p}{\alpha'_p},
\end{equation} 
with  
\begin{equation}
\alpha'_p = 1 - \sum_c \left[\tfrac{1}{2} v_{cp} \rho_{c} - \tfrac{3}{8} v_{cp}^2 \rho_c^2 \right].
\end{equation}  

Similarly, the presence of several colloidal species reduces the effective available volume to the individual colloidal species. Namely, for species $c$, we have

\begin{equation}
\alpha'_c = 1 - \sum_{d \neq c} \left[\tfrac{1}{2} v_{cd} \rho_d - \tfrac{3}{8} v_{cd}^2 \rho_d^2 \right],
\end{equation} 
such that $c,d \in \{\rm n+1,...,\rm n+m\}$. Note that with the current implementation we do not consider attraction between colloidal species, and thus $\alpha'_c$ is irrelevant for this work, but valid in general.

In terms of these effective densities, the mean-field contribution to the bulk free energy due to polymer-polymer and polymer-colloids interaction is given by the enhanced mean field functional
\begin{equation}
\begin{split}
\beta \mathcal{F}_{\rm EMF} = \tfrac{1}{2} \sum_{i,j} \alpha'_{i}\,\rho'_i \rho'_j \,\hat{V}_{ij}.
\label{eq.meanfield}
\end{split}
\end{equation}
Note that the standard mean-field (SMF) is obtained by setting $\alpha'_i=1,\>\forall i$, and replacing the effective densities with their original values, i.e.
\begin{equation}
\begin{split}
\beta \mathcal{F}_{\rm SMF} = \tfrac{1}{2} \sum_{i,j} \rho_i \rho_j \,\hat{V}_{ij}  ,
\label{eq.meanfield}
\end{split}
\end{equation}
where $\hat{V}_{ij}$ are the integrated strength \cite{PhysRevE.64.041501}, and
where $i,j \in \{1,\ldots,\rm n+m\}$. The factors $\alpha'_{i}$ ensure the correct normalization of particle numbers when integrating over the total volume of the system~\cite{Lekkerkerker_1992}. 

Here, the integrated strength for any pair of species $i$ and $j$, is defined as \cite{PhysRevE.64.041501, brader2003statistical} 
\begin{equation}
\beta \hat{V}_{ij} = 4\pi\int r^2\phi_{ij}(r)dr,
\label{eq::fourier}
\end{equation}
which, for a pair interaction defined by Eq.~(\ref{eq::pol-pol}), can be expressed analytically as $\hat{V}_{pq} = \pi^{3/2} \, \epsilon_{pq} R_{pq}^3 $. 
%When it comes to the attractive contribution, we assume that the attractive range in Eq.~(\ref{eq:coll-poll_att}) extends down to $r=0$ when performing the integration~\textcolor{red}{[Vikki cites relevant work]}.
%By doing so, we get  $\hat{V}_{cp} = \frac{4\pi}{3} \epsilon_{cp} \delta^3\sigma_{cp}^3$. 
To simplify the notation, we will henceforth use $\hat V$ to represent the dimensionless form, $\beta \hat V \sigma^{-3}$, where $\sigma$ defines the unit of length.

%The integrated strength of the attractive contribution of the colloid-polymer interaction takes the form 
%\begin{widetext}
%\begin{equation}
%\hat{V}_{pc} = -4\pi\dfrac{\epsilon}{3\lambda} r_{\rm min}^3 +\frac{\rm u}{\rm u-v} \left(\frac{\rm u}{\rm v}\right)^{\rm \!v/(u-v)}
% 4 \pi\varepsilon \sigma_{ac}^3 \Big[ \lambda_p \frac{1}{\rm u-3} \left(\frac{\sigma_{ac}}{r_{\rm min}}\right)^{\rm u-3} - \frac{1}{\rm v-3} \left(\frac{\sigma_{ac}}{r_{\rm min}}\right)^{\rm v-3} \Big],
%\label{eq::fourier2}
%\end{equation}
%\end{widetext}
%\textcolor{red}{Still problem with variables. Different $\varepsilon/\epsilon$, what is $\lambda_p$, what is %$\sigma_{ac}$?}
%where $r_{\rm min}=(2\lambda)^{\rm 1/ \rm (u-v)}\sigma_{cp}$ is the minimum of Eq.~(\ref{eq::coll-pol_full}).

%%%%%%%%%%%%%%%%%%%%%%%%%%%%%%%%%%%%%%%%%%%%%%%%%%%%%%%%%%%%%%%%%%%%%%%%%%%%%%%%%%%%%%%%%%%%%%%%%%%%%%%%%%%%%%%%%%%%%%%

\subsection{Coexistence criteria}\label{sec::coexistence}
The chemical potential for any species $i$ can be expressed as
\begin{equation}
\mu_i = -\frac{\partial \mathcal{F}}{\partial \rho_i}, 
\label{eq.4_17}
\end{equation}
where $\mathcal{F}$ %\textcolor{violet}{Finally Checked.....!!!!! its all correct there is no doubt left I am damn sure}
is the Helmholtz free energy per unit volume given in Eq.~(\ref{eq::total_free_energy}). On the other hand, the total pressure, in the absence of any external potential, can be written as% \cite{D4SM00332B}
\begin{equation}
    P = -\mathcal{F}  + \sum_{i} \mu_i \rho_i,
\label{eq.4_18}
\end{equation}
where the summation runs over all species.

%\textcolor{magenta}{Now I think no need to ask professor Andy about his own paper on Ouzo where he has given the same formula for the pressure I am citing him in this section it would be better I think. We should also note that the formula is given in per unit volume there also as we are giving...}

%\textcolor{cyan}{I have also removed the $\rho_i$ from pressure as in case of Prof Andy's work. But we did not cite his ternary mixture work he might not be feeling happy about that... really regretting... it didn't come in mind also mine...}

Let us consider the existence of two coexisting phases, phase-I and phase-II. Following the Gibbs' criteria, the coexisting conditions are
\begin{equation}
\mu_i^{\rm I} = \mu_i^{\rm II}\qquad {\rm and} \qquad P^{\rm I} = P^{\rm II},
\label{eq.4_26}
\end{equation}
where $i \in \rm \{1,\ldots,n+m \}$. Under these constraints, we are left with $\rm n+m+1$ equations to solve for $\rm 2(n+m)$ variables, representing the coexisting densities of all species in the two phases. We thus have an undetermined system of equations. To solve this, we need to add additional constraints. To do so, let us fix the density of each species such that 
\begin{equation}
V_{\rm f}\rho_i^{\rm I}  + (1-V_{\rm f})\rho_i^{\rm II} = \bar{\rho}_i,
\label{eq::total_density_1}
\end{equation}
with
\begin{equation}
\bar{\rho}_i = N_i/V,
\label{eq.4_29}
\end{equation}
where $N_i$ is the total number of particles of species $i$, $V$ is the total volume of the system, and $\bar\rho_i$ the average density of species $i$. $V_{\rm f}$ is the volume fraction representing phase-I, thus $(1-V_{\rm f})$ represents the volume fraction of phase-II. By doing so we add an extra variable, $V_{\rm f}$, which provides $\rm n+m$ new equations through (\ref{eq::total_density_1}). We are now left with $\rm 2(n+m)+1$ equations and $\rm 2(n+m)+1$ variables. The same scheme can be readily extended in the case of arbitrary many coexisting phases. 

Solving the coexistence equations in the current formulation is challenging, as the wide range of potential coexisting densities for many species hinders numerical convergence. To simplify the approach, we express the coexisting densities as fraction of total densities, such that

\begin{equation}
\rho_i = \rho_{\rm t} \times x_{i-1} \prod_{j=i}^{\rm n+m-1} x'_j    
\label{eq::relative_densities}
\end{equation}
where $\rho_{\rm t} = \sum_i\rho_i$, $i \in \{\rm 1,...,n+m\}$,  $x_i \in [0,1 ], $ $x'_{i} = {(1-x_i)}$ and $x_{0}, x'_{0} =1$. On the other hand, the reduced densities can be expressed as
\begin{equation}
x_i = \frac{\rho_{i+1}}{\sum_{j=1} ^{i+1} \rho_j}.
\label{eq.4_31}
\end{equation}
This nested construction ensures that all species densities remain positive.
The structure can be interpreted as a sequence of branching ratios, where each $x_i$ governs the split of remaining density among successive species. This representation is particularly well-suited for numerical optimization and phase coexistence solvers, as it transforms the constrained problem of non-negative, normalized densities into an unconstrained optimization over $\rho_{\rm t}$ and $\{x_i\} \in (0,1)$. The proposed reformulation helps in restricting the solver to find a disjoint solution using simple numerical solving modules, such as, e.g., \texttt{hybr}, freely available in python. The numerical approach consists in the following: one begins by generating a trial solution in terms of reduced density. This trial solution is then transformed into real densities, which are substituted into the coexistence equations. The values are iteratively updated until convergence is achieved toward a disjoint solution in the reduced variables. Once convergence is confirmed, the solution is converted back into real densities. A detailed description of these steps is provided in Section II of the supplementary information (SI).

%Sometimes it is also feasible to change the relevant functions in the form of reduced densities. Especially, if the free energy functional $f$ has simple form, e.g, in case if all the species are polymers.  In that case we would not need to change the densities in the real densities to get the value of equation at the particular point in the density space.  

Using this reduced-densities formulation, the chemical potential for the $i^{\rm th}$ component can be written as
\begin{equation}
\begin{split}
\mu_i = 
f +\rho_{\rm t} \frac{\partial f}{\partial \rho_{\rm t}} \\ & \hspace{-2cm} + \sum_{\substack{S \subseteq \{x_{i-1}, \dots, x_{\rm n+m-1}\} \\ S \ne \emptyset}}   
\left( 
\frac{\partial^{|S|} f}
{\prod\limits_{x_j \in S} \partial x_j}
\prod\limits_{x_j \in S} w_j (x_j)
\right),
\end{split}
\label{eq.4_32}
\end{equation}
where free energy per particle, $f$, is expressed in terms of reduced densities ($\{\rho_{\rm t}, x_1,...,x_{\rm n+m-1} \}$ for an $\rm n+m$-component system). $\substack{S \subseteq \{x_{i-1}, \dots, x_{\rm n+m-1}\}}$ represents summation performed over all the possible subsets (all possible combination of different number of species and their permutation). However all the elements must belong to the set $\{x_{i-1},...,x_{\rm n+m-1} \}$.  $w_j({x_j})$ is given by
\begin{equation}
w_j(x_j) = 
\begin{cases}
1 - x_{j}, & \text{if } j = i \text{ and } i > 1 ,\\[4pt]
- x_j, & \text{otherwise}.
\end{cases}
\label{eq.4_33}
\end{equation}
Within this formulation, the pressure now reads
\begin{equation}
P = \rho_{\rm t}^2 \frac{\partial f}{\partial \rho_{\rm t}}.
\label{eq.4_34}
\end{equation}

%%%%%%%%%%%%%%%%%%%%%%%%%%%%%%%%%%%%%%%%%%%%%%%%%%%%%%%%%%%%%%%%%%%%%%%%%%%%%
%%%%%%%%%%%%%%%%%%%%%%%%%%%%%%%%%%%%%%%%%%%%%%%%%%%%%%%%%%%%%%%%%%%%%%%%%%%%%
\subsection{Grand potential for inhomogeneous system}
The grand canonical potential of an inhomogeneous system, $\Omega$, can be written as \cite{Evans01041979}

\begin{equation}
\begin{split}
\Omega[\{\rho_i(\textbf{r})\}] &=  F_{\rm id}^{\rm in}[\{\rho_i(\textbf{r})\}] + F_{\rm exc}^{\rm in}[\{\rho_i(\textbf{r})\}] \\&+ \sum_i\int \left( V_i^{\mathrm{\rm ext}}(\mathbf{r}) - \mu_{i} \right) \, \rho_i(\textbf{r})d\mathbf{r},
\label{eq.4_25}
\end{split}
\end{equation}
where $V_i^{\mathrm{\rm ext}}(\mathbf{r})$ represent any external scalar potential and $\mu_i$ is the chemical potential of species $i \in \{1,..., \rm n+m\}$, which is defined in bulk. For inhomogeneous systems, the (total) ideal free energy contribution reads
\begin{equation}
    \beta F^{\rm in}_{\rm id}[\{\rho_i(\textbf{r})\}]=\sum_i\int\rho_i(\textbf{r})\left[\ln(\Lambda_i^3\rho_i(\textbf{r}))-1\right]d\textbf{r}.
\end{equation}
The (total) excess free energy is split into two terms, namely

\begin{equation}
\begin{split}
F_{\rm exc}^{\rm in}[\{\rho_i(\textbf{r})\}] = \int \left ( \mathcal{F}_{\rm AOV}^{\rm in}[\{\rho_i(\mathbf{r})\}]+\mathcal{F}_{\rm EMF}^{\rm in}[\{\rho_i(\mathbf{r})\}] \right) d \mathbf{r}.
\label{eq.excess_free_energy}
\end{split}
\end{equation}

%\textcolor{violet}{Checked..........!!!!!!!!!!}

The colloid-polymer system is described by means of the cavity model within the Rosenfeld formalism, which is compatible with dimensional crossover, and is given by \cite{PhysRevLett.85.1934}
\begin{equation}
\beta\mathcal{F}_{\rm{AOV}}^{\rm in} = \Phi_1 +\Phi_2 + \Phi_3.
\label{eq.4_19}
\end{equation}
The individual terms are defined as
\begin{equation}
\Phi_1 = \sum_{i=1}^{\rm n+m} [n_0^i \frac{\partial \phi_0}{\partial n_3^i}],
\label{eq.4_20}
\end{equation}
\begin{equation}
\Phi_2 = \sum_{i=1,j=1}^{\rm n+m}\left(n_1^in_2^j- \mathbf{n}_{\rm V1}^i \cdot \mathbf{n}_{\rm V2}^j  \right)\frac{\partial^2 \phi_0}{\partial n_3^i \partial n_3^j}
\label{eq.4_21}
\end{equation}
and
\begin{widetext}
\begin{equation}
\Phi_3 =
\frac{1}{8\pi}
\sum_{\substack{i=1,j=1\\k=1}}^{\rm n+m}
\Bigg(
   \frac{n_2^i n_2^j n_2^k}{3} - n_2^i \, \mathbf{n}_{\rm V2}^j \cdot \mathbf{n}_{\rm V2}^k  +
   \frac{3}{2} \left( \mathbf{n}_{\rm V2}^i  \mathcal{T}^{j} \mathbf{n}_{\rm V2}^k 
   - \mathrm{tr} \left[\mathcal{T}^i \mathcal{T}^j \mathcal{T}^k \right] \right) \Bigg)
  \frac{\partial^3 \phi_0}{\partial n_3^i \partial n_3^j \partial n_3^k},
\label{eq.4_22}
\end{equation}  
\end{widetext}
where $\phi_0$ represents the zero-dimensional free energy ($\beta \mathcal F_0$) given by $\beta \mathcal{F}_0 = (1 -\sum_{i=\rm n+1}^{\rm n+m} \eta_i - \sum_{j=1}^{\rm n} \eta_j)\ln(1- \sum_{i=\rm n+1}^{\rm n+m} \eta_i) - \sum_{i=\rm n+1}^{\rm n+m} \eta_i$. $n_0^i, n_1^i, n_2^i, n_3^i, \mathbf{n}_{\rm V1}^i, \mathbf{n}_{\rm V2}^i$ and $\mathcal{T}$ are the weighted densities, which, for inhomogeneous cases, are given by
\begin{equation}
    n_k^i(\mathbf{r}) = \int \omega_k^i(\mathbf{r'-\mathbf{r}}) \rho_i(\mathbf{r'}) d\mathbf{r'}, \quad k=0,\dots,3,  \rm V1, V2,
    \label{eq.wdensity}
\end{equation}
where $\omega_k^i$ represent weights functions. The the scaler weight functions read
\begin{equation}
    \begin{aligned}
        \omega_0^i(\mathbf{r}) &=\frac{1}{4\pi R_{i}^2} \delta(\mathbf{r}- R_i),\\
        \omega_1^i(\mathbf{r}) &=\frac{1}{4\pi R_{i}} \delta(\mathbf{r} - R_i),\\
        \omega_2^i(\mathbf{r}) &= \delta(\mathbf{r}-R_c),\\
        \omega_3^i(\mathbf{r}) &= \Theta(\mathbf{r}-R_c).
    \end{aligned}
\label{eq.weights}
\end{equation}
Here $R_i$ is the radius of the colloids, i.e. $R_i=\sigma_{ii}/2$, $\Theta (\mathbf{r} - R_i)$ is the Heaviside step function, and $\delta(\mathbf{r}-R_i)$ is the Dirac-delta function. The vector and tensor weight functions are defined as $\mathbf{\omega}_{\rm V1}^i (\mathbf{r}) = \omega_1^i(\mathbf{r}) \mathbf{r}/r$, $\mathbf{\omega}_{\rm V2}^i(\mathbf{r}) = \omega_2^i (\mathbf{r})\mathbf{r}/r$  and  $\omega_\mathcal{T}^i(\mathbf{r}) = \omega_2^i (\mathbf{r})[\mathbf{r}\mathbf{r}/r^2 - \mathcal{I}/3] $, where tr is the classical trace, $\rm {tr} (\omega_{\mathcal{T}}^i) = 0$ and $\mathcal{I}$ is the 3x3 identity matrix. The corresponding weighted densities are obtained in the same way as in Eq.~(\ref{eq.wdensity}), for both colloids and polymers. The details of the methodology and its application is given in the SI. %\textcolor{green}{check later}

Note that the AOV model, which accounts for the inhomogeneity of the system, provides a good approximation of the depletion effects in colloidal suspensions \cite{PhysRevLett.85.1934}. In bulk, this formulation reduces to the Rosenfeld formalism for hardcore interactions combined with FV theory, which captures the polymer-induced depletion forces, as discussed earlier.

In the inhomogeneous case, the EMF contribution in $F^{\rm in}_{\rm exc}$ is given by
\begin{equation}
\begin{split}
\hspace{0.5cm}\beta \mathcal{F}_{\rm EMF}^{\rm in}[\{\rho_i(\mathbf{r})\}] &=\frac{1}{2} \sum_{i,j} \rho_i(\mathbf{r}) \\ & \times
\int \frac{\rho_j(\mathbf{r}')}{\alpha'_j[\{ \rho_c(\mathbf{r}') \}]} \,\phi_{ij}(|\mathbf{r}-\mathbf{r}'|) d\mathbf{r}' .
\label{4.16}
\end{split}
\end{equation}
%\textcolor{violet}{Checked!!!!!!!! its per unit volume functional basically ...........!!!!!!!!}
Here, $i,j \in \{1, ...,\rm n+m \}$ and $c \in \{\rm n+1,...,n+m \}$, and $\phi_{ij}$ represents both the polymer-polymer repulsive potential via Eq.~(\ref{eq::pol-pol}) and the colloid-polymer attraction. For the latter, instead of using the square well form in Eq.~(\ref{eq:coll-poll_att}), we use a parametrized Gaussian potential, as detailed in the SI, which by being continuous provides a better choice for obtaining the numerical solutions of the inhomogeneous densities in one dimension.

The inhomogeneous equilibrium density profile are obtained by minimizing the grand potential $\Omega$ with respect to the densities of the different species, providing the solution \cite{Roth_2010}
\begin{equation}
\begin{split}
    \rho_i(\mathbf{r}) = e^{[-V_i^{\rm ext}(\mathbf{r}) + c^{(1)}_i(\mathbf{r}) + \beta \mu_i]},
\end{split}\label{eq::inhomog_density}
\end{equation}
where $c^{(1)}_i(\mathbf{r}) = -\beta  \frac{\partial \mathcal{F}_{\rm exc}^{\rm in}[\{ \rho_i\}]}{\partial \rho_i(\mathbf{r})}$, is the one body direct correlation function.

\section{Results for a ternary mixture with two polymers and one colloidal species}

\subsection{Bulk phase properties}

\subsubsection*{Setup and free energy landscape}

\begin{figure}[H]\centering
\includegraphics[width=\linewidth]{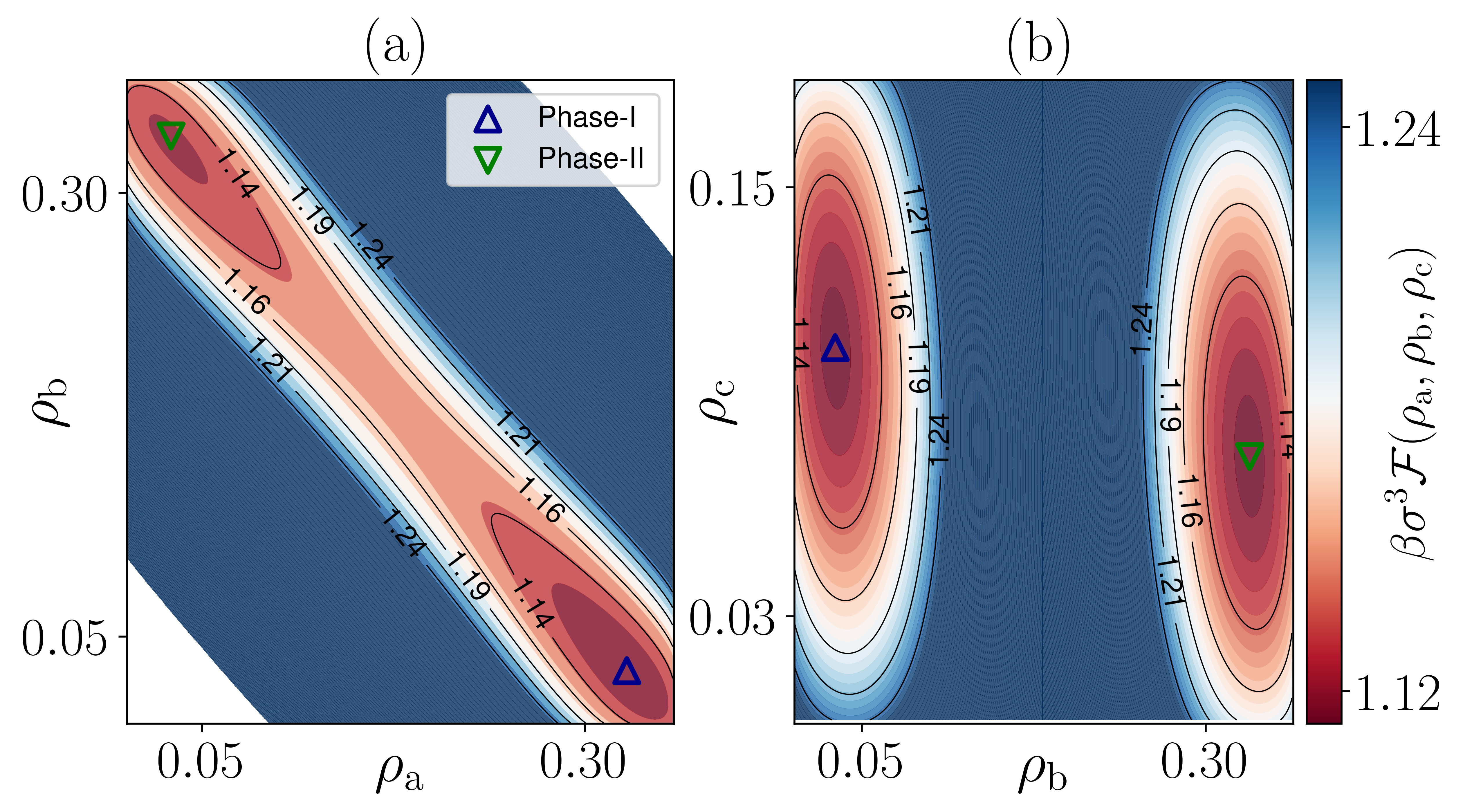}
  \caption{
(a) Free energy density for a given pair of coexisting densities in the \(\rho_{\rm a} \rho_{\rm b}\)-plane and  
(b) in the \(\rho_{\rm b} \rho_{\rm c}\)-plane, where a and b are two immiscible polymer species and c a colloidal species.  
The upper and lower triangles correspond to phase-I and phase-II, respectively. The color bar indicates the value of $f$ at each point in the density space, where the two conjugate points represent the same system. All conjugate points represent the same canonical ensemble with  
\(\bar{\rho}_{\rm a} = \bar{\rho}_{\rm b} = 0.18\) and \(\bar{\rho}_{\rm c} = 0.09\).  
%The colloids have identical sizes with radius of gyration \(R_{aa} = R_{ab} = R_{bb} = 1.414\),  
%interaction strengths \(\epsilon_{aa} = \epsilon_{bb} = 2.0\), and \(\epsilon_{ab} = 2.5\).  
%The colloid size is \(\sigma_{cc} = 1.0\), while polymers have excluded-volume sizes  
%\(\sigma_{aa} = \sigma_{bb} = 1.0\).  
%Additionally, polymer \(a\) and the colloid interact via an attractive potential with integrated strength  
%\(\hat{V}_{ac} = -2.0\). While polymer $b$ and colloid have hardcore repulsion. 
}
\label{fig.4_2}
\end{figure}

To provide a concrete example of our general approach, we focus on a binary mixture of immiscible polymers, a and b, in which a single colloidal species, c, is dispersed. This situation is key in water purification by means of ATPSs  \cite{iqbal2016aqueous}. With $\sigma$ defining the length scale of the system, we initially focus on the symmetric case where $R_{\rm aa} = R_{\rm ab}= R_{\rm bb} = \sqrt{2} \sigma$. Without loss of generality, we set $\sigma=1$ here. The repulsion strength between the polymers are set to correspond to polymers in good solvent, i.e. $\epsilon_{\rm aa} = \epsilon_{\rm bb} =2.0$  \cite{10.1063/1.1344606}, while the cross repulsion parameter is set to $\epsilon_{\rm ab}=2.5$. This setup fulfills the condition for phase separation at fixed volume, given by \cite{PhysRevE.62.7961}
\begin{equation}
\chi = 2\hat{V}_{\rm ab}  - [ \hat{V}_{\rm aa} + \hat{V}_{\rm bb}] > 0.
\label{eq::phase_sep}
\end{equation}
For modeling the excluded volume arising from the interaction between polymers and colloids (relevant in $\mathcal{F}_{\rm HC}$ and $\mathcal{F}_{\rm FV}$) we set here $\sigma_{\rm aa} = \sigma_{\rm bb} = \sigma_{\rm cc}=\sigma$. For simplicity, we set some attraction between polymer a and the colloids such that $\hat{V}_{\rm ac} = -2.0 $, while the interaction between polymer b and the colloid, as well as the colloid-colloid, are purely repulsive, i.e. $\hat{V}_{\rm cc} , \hat{V}_{\rm bc} = 0$. Note that, in bulk, specifying the values of $\delta$ and $\epsilon_{cp}$ of Eq.~(\ref{eq:coll-poll_att}) is not necessary, due to the integration in Eq.~(\ref{eq::fourier}). Also, since all sizes are the same here, the only driving force for colloids partitioning rise from their asymmetric attraction with the two polymer species.

We consider the existence of only two phases, with 
\(\{ (\rho_{\rm a}, \rho_{\rm b}, \rho_{\rm c})^{\mathrm{I}}, (\rho_{\rm a}, \rho_{\rm b}, \rho_{\rm c})^{\mathrm{II}} \}\), as in Eq.~(\ref{eq::total_density_1}). These densities are obtained by minimizing the free energy defined by Eqs.~(\ref{eq::total_free_energy}), (\ref{eq::ideal_free_energy}), (\ref{eq.rosenfeld}), (\ref{eq.freevolume}) and (\ref{eq.meanfield}), under the conditions specified so far. To simplify notation, we will henceforth use $\rho$ to represent the dimensionless form, $\sigma^{3}\rho$, for all densities. In Fig.~\ref{fig.4_2}, we show an example of free energy landscape for the case $\bar{\rho}_{\rm a} = \bar{\rho}_{\rm b} = 0.18$ and $\bar{\rho}_{\rm c} = 0.09$. As expected, we can see that the energy landscape converges into two well-separated phases, noted as phase-I and phase-II, respectively. This can be seen in both the $(\rho_{\rm a} \rho_{\rm b})$-plane (panel (a)) and  
in the \(\rho_{\rm b} \rho_{\rm c}\)-plane (panel (b)).

\subsubsection*{Bulk phase response: Comparison with simulations}
\begin{figure}[h!]
\includegraphics[width=\linewidth]{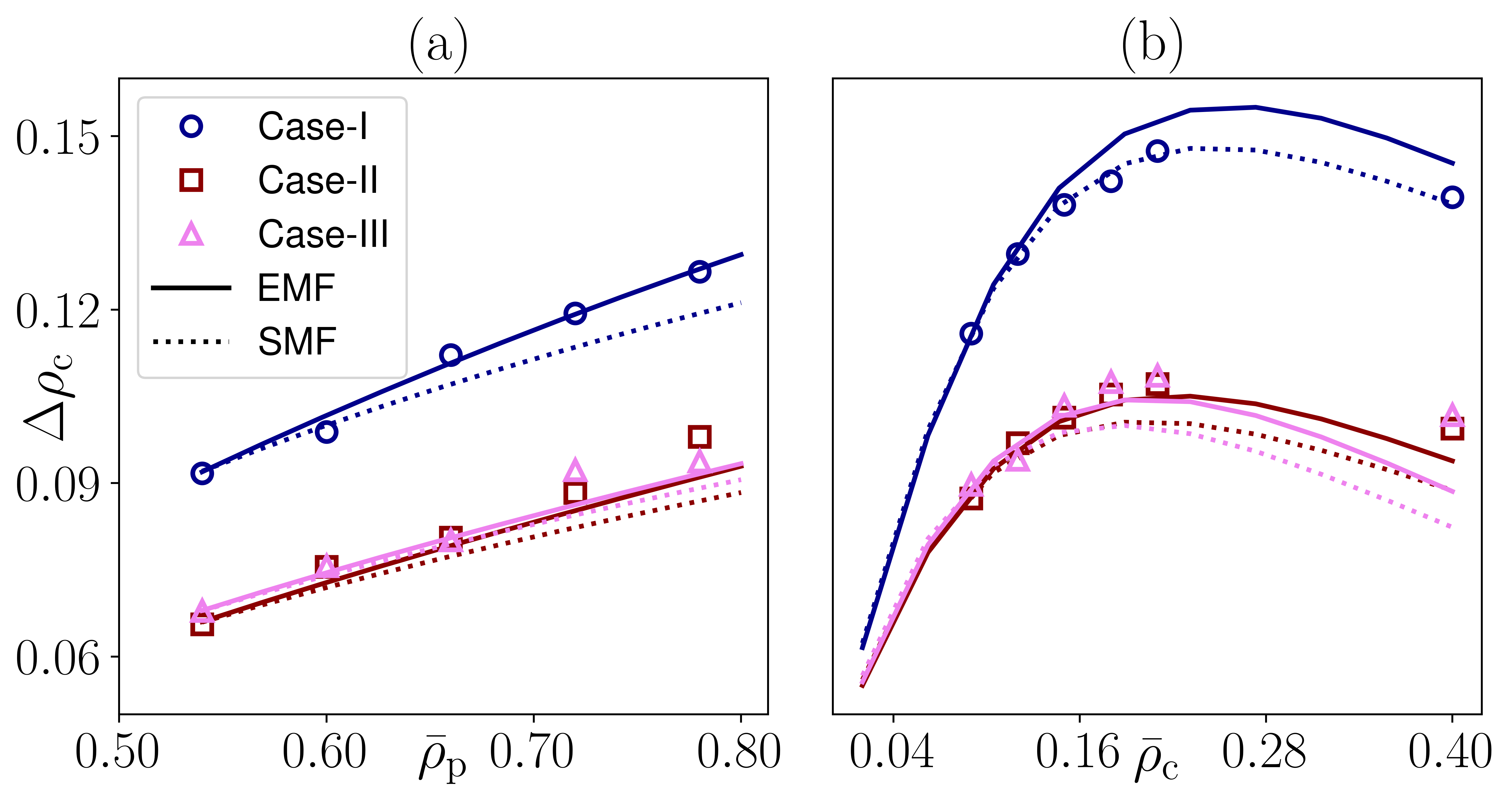}
  \caption{Density difference between the two phases of the colloidal species, $\Delta \rho_{\rm c} = \rho_{\rm c}^{\rm I} - \rho_{\rm c}^{\rm II}$ are shown using symbols for particle-based Brownian dynamics simulations (similar to those in \cite{SCACCHI20251135}). The simulation results, for the parameters summarized in Table~\ref{tab:interaction}, are shown for case-I, case-II and case-III, and are compared with our cDFT using EMF (solid lines), as well as SMF (dashed lines). In panel (a) we show the results while varying the total polymer concentration ($\bar\rho_{\rm p} = \bar\rho_{\rm a} +\bar\rho_{b\rm }$) at fixed colloid concentration $\bar\rho_{\rm c} = 0.09$. In panel (b), the results are shown for varying colloid density at fixed polymer concentration $\bar\rho_{\rm a} = \bar\rho_{\rm b} = 0.27$.  
}
\label{fig.phase_response}
\end{figure}

Before investigating the effect on phase separation of different physical properties of this ternary system, we provide a comparison between cDFT and particle-based simulations (similar to those in Ref.~\citenum{SCACCHI20251135}). Specifically, we set the polymer-polymer interaction to be gaussian potential, as in Eq.~({\ref{eq::pol-pol}}). For colloid-colloid and colloid-polymer b, we use a standard Lennard-Jones potential cut at its minimum, to simulate purely repulsive interactions. On the other hand, for the colloid-polymer a interaction, we consider two different Mie potentials with exponents u and v. Specifically, we use $48-24$, $24-12$ and $12-6$ potentials. For completeness, the $\rm u-v$ Mie potential is given as
\begin{equation}
\phi_{\rm ac}^{\rm sim}(r) =
\dfrac{\rm u}{\rm u - v} \left(\dfrac{\rm u}{\rm v}\right)^{\rm v/(u-v)} \epsilon_{\rm ac}
\Bigg[\left(\frac{\sigma_{\rm ac}}{r}\right)^{\rm u} - \left(\frac{\sigma_{\rm ac}}{r}\right)^{\rm v} \Bigg].
\label{eq::coll-pol_simulation}
\end{equation}
The parameters used for these interactions are summarised in Table \ref{tab:interaction}. The simulations are run in a periodic box with size $15\sigma \times15\sigma\times60\sigma$ for a total of $2\times10^8$ steps, of which the first half is considered as equilibration time. The dimensionless time step is set to $d\tau=10^{-4}\sigma^2D^{-1}$, where $D$ is the particle self-diffusion constant. Equilibration is assessed by observing the individual density profiles until they show no further changes.

\begin{table}[h]
\centering
% First block: Polymer–Polymer
\begin{tabular}{|c|c|c|c|}
\hline
\multicolumn{4}{|c|}{Polymer–Polymer - Eq. (\ref{eq::pol-pol}) \hspace{2cm}} \\ \hline
Species & $R_{pq}$  & $\epsilon_{ij}$ & $r_{\rm cutoff}$\\ \hline
aa & $\sqrt{2}$ & $2.0$ & 3.5 \\
bb & $\sqrt{2}$ & $2.0$ & 3.5\\
ab & $\sqrt{2}$ & $2.5$ & 3.5\\ \hline
\end{tabular}

\vspace{0.2cm} % vertical space between the two blocks

% Second block: Colloid–Polymer
\begin{tabular}{|c|c|c|c|c|c|}
\hline
\multicolumn{6}{|c|}{Colloid-Colloid and Colloid–Polymer - Eq. (\ref{eq::coll-pol_simulation})\hspace{2cm}} \\ \hline
Species & u & v & $\epsilon_{ij}$ & $\sigma_{ij}$ & $r_{\rm cutoff}$ \\ \hline
cc       & 12 & 6  & $1$      & $1$ & $2^{1/6}$\\
bc       & 12 & 6  & $1$      & $1$ & $2^{1/6}$ \\
ac (case-I)   & 12 & 6  & $0.1757$ & $1$ & 5 \\
ac (case-II)  & 24 & 12 & $0.3104$ & $1$ & 5 \\
ac (case-III)  & 48 & 24 & $0.7565$ & $1$ & 5 \\ \hline
\end{tabular}

\caption{Interactions between all system components in particle-based simulation in bulk.}
\label{tab:interaction}
\end{table}

Also in the corresponding cDFT system, polymer-polymer interactions are defined as in Eq.~(\ref{eq::pol-pol}), and their parameters as for particle-based simulations (see top part of Table~\ref{tab:interaction}). When it comes to describing colloid–colloid and colloid–polymer interactions, a one-to-one benchmark is not straightforward. In fact, in our particle-based simulations these interactions are soft, while cDFT considers hard-core repulsion plus eventually attraction. To solve this challenge, we set the size of the corresponding hard-core in cDFT following the Barker-Henderson \cite{10.1063/1.1701689} (BH) perturbation theory. BH theory provides a mapping from a soft repulsive potential to a hard-core diameter, $\sigma_{\rm BH}$, which is given by
\begin{equation}
\sigma^{\rm BH}_{ij} = \int_0^{\sigma_{ij}} \Big[ 1 - e^{-\beta \phi_{ij}(r)} \Big] dr,
\label{eq::bh_radius}
\end{equation}
where $\phi_{ij}$ correspond to the potentials in particle-based simulations with parameters given in the bottom part of Table~\ref{tab:interaction}. From these interaction potentials and parameters, we obtain $\sigma^{\rm BH}_{\rm cc}=\sigma^{\rm BH}_{\rm bc}=0.973$ and $\sigma^{\rm BH}_{\rm ac}=0.874$ for case-I, $0.969$ for case-II and 1.01 for case-III, respectively. We are now left with defining the corresponding attraction strength between polymer a and colloid. To achieve this, we calibrate the value of $\hat{V}_{\rm ac}$ such that $\Delta \rho_{\rm c} = \rho_{\rm c}^{\rm I} - \rho_{\rm c}^{\rm II}$, i.e. the difference in coexisting densities of colloids, in cDFT matches the corresponding particle-based system. We do this under the following conditions: $\bar{\rho}_{\rm a} = \bar{\rho}_{\rm b} =0.27$, $\bar{\rho}_{\rm c} = 0.09$. For case-I (see Table~\ref{tab:interaction}), we find a good match for $\hat{V}_{\rm ac} = -3.0$, while for case-II for $\hat{V}_{\rm ac} = -2.8$, and for case-III for $\hat{V}_{\rm ac} = -2.5$.

To show the validity of the cDFT predictions on the phase response of the system, in Fig.~\ref{fig.phase_response}(a) we vary the total polymer density, $\bar{\rho}_{\rm p}$, from 0.54 to 0.8 with $\bar{\rho}_{\rm a} = \bar{\rho}_{\rm b}$, while keeping $\bar{\rho}_{\rm c}$ fixed at $0.09$. The predictions of our cDFT are bench-marked against particle-based simulations for all the three cases. Panel (a) shows a monotonic increase in partitioning as the total polymer density is increased. We also see that our improved MF approach performs better in all cases up to high colloidal density. Similarly, in panel (b) of Fig.~\ref{fig.phase_response}, we vary the density of the colloids from $\bar{\rho}_{\rm c} = 0.09 $ to $0.4$ while keeping $\bar{\rho}_{\rm a} = \bar{\rho}_{\rm b} = 0.27$ fixed. Interestingly, we see a decrease in partitioning after reaching a maximum value at around $\bar{\rho}_{\rm c} =  0.2$. The presence of colloids increases the effective polymer density by reducing the available volume, which in turn lowers the free energy through enhanced polymer–colloid attraction. This effect, however, reaches a maximum, as the gain is compensated by increased colloid–colloid and polymer–polymer repulsive interactions, along with a further reduction in available volume that limits polymer–colloid contributions when colloids are too close to each other. Also here, the cDFT predictions are in good agreement with simulations, up to high colloidal concentrations. We can see that, with the exception of case-I, the EMF approach performs better than its standard version. To investigate this specific case, the sensitivity of both MF approaches with respect to the value of $\sigma_{\rm BH}$ has been addressed. We find that the EMF approach remains more anchored to the presented predictions, while the SMF approach deviates from the presented curve quite rapidly. All-in-all, we can conclude that the EMF free energy functional proposed here (Eq.~(\ref{eq.meanfield})) performs better than its standard version.

\subsubsection*{Effect of polymer properties on the phase response}

\begin{figure}[h!]
\includegraphics[width=\linewidth]{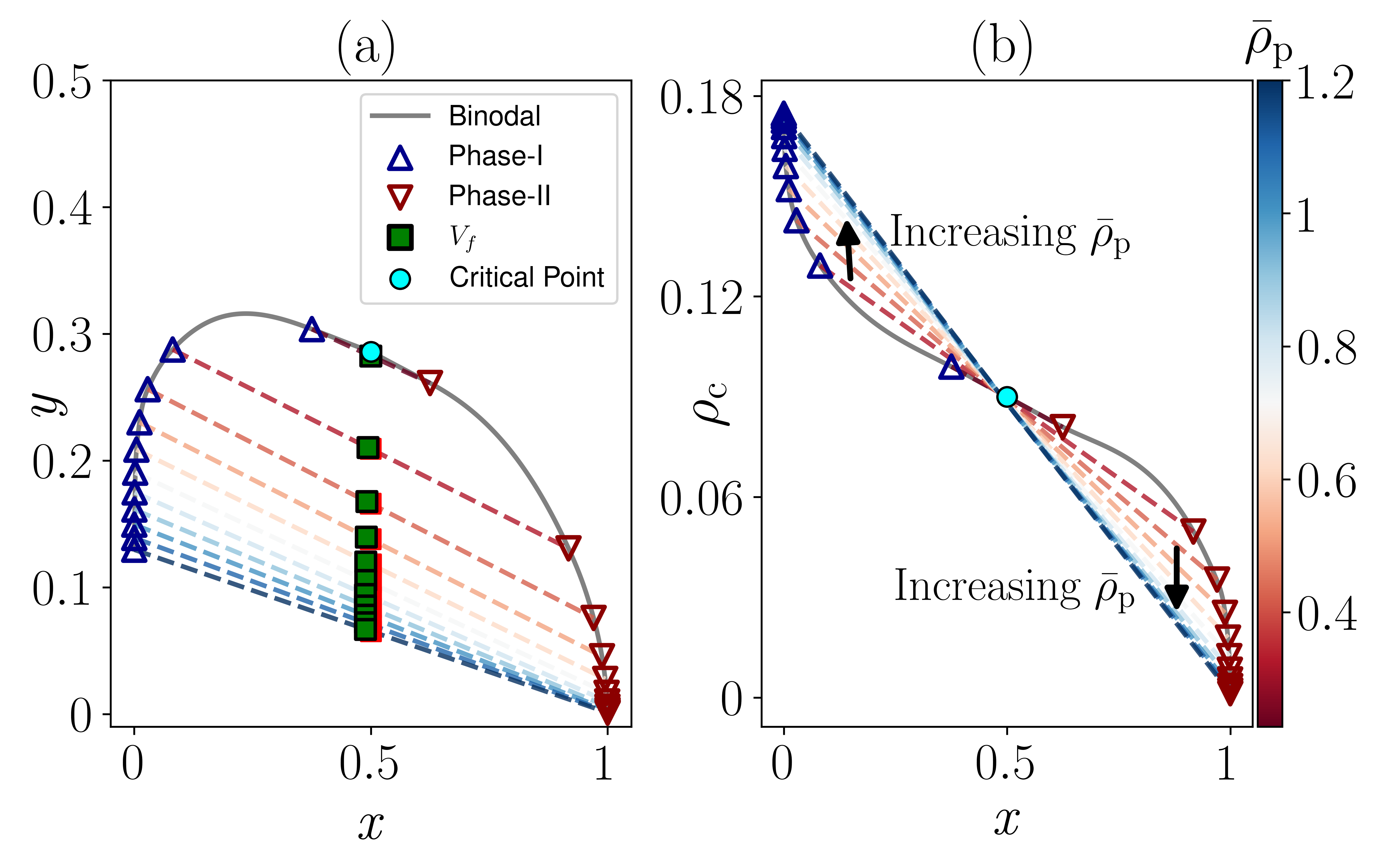}
  \caption{In (a), binodal curve for the ternary colloid-polymer mixture described by the following parameters: $R_{\rm aa} = R_{\rm bb} = R_{\rm ab} =  \sqrt{2} \sigma$, $\epsilon_{\rm aa} = \epsilon_{\rm bb} = 2.0, \epsilon_{\rm ab} =  2.5$, $\sigma_{\rm aa} = \sigma_{\rm bb} = \sigma_{\rm cc}=1$ and $\hat{V}_{\rm ac}=-5.6$. $x$ is the fraction of polymer b with reference to the total polymer densities $\bar\rho_{\rm p}$ and $y$ is the fraction of colloids with reference to the total density $\bar\rho_{\rm t}$. Here $\bar\rho_{\rm c} = 0.09$ is kept fixed. The upper triangles indicate phase-I (polymer a-rich phase), while the lower triangles indicate phase-II (polymer b-rich phase). Coexisting points are connected by isobars, where their color corresponds to the value of $\bar\rho_{\rm p}$. The green squares represent the value of $V_{\rm f}$, i.e. the volume ratio of phase-I, while the red squares represent $V_{\rm f}=0.5$. The full circle represents the critical point. In (b), effect of polymer concentration on the coexistence densities of colloids.}
\label{fig.phase_diagram}
\end{figure}

In order to further explore the effect of polymers properties on the phase response of the ternary mixture, we calculate the coexisting densities for a system at different total polymer concentration $\bar{\rho}_{\rm p} = \bar{\rho}_a + \bar{\rho}_b$. We explore $\bar{\rho}_{\rm p} $ ranging between 0.23 and 1.1 at fixed average colloid concentration $\bar{\rho}_{\rm c }= 0.09$. Here the parameters are as follow: $R_{\rm aa} = R_{\rm bb} = R_{\rm ab} =  \sqrt{2} \sigma$, $\epsilon_{\rm aa} = \epsilon_{\rm bb} = 2.0, \epsilon_{\rm ab} =  2.5$, $\sigma_{\rm aa} = \sigma_{\rm bb} = \sigma_{\rm cc}=\sigma$ and $\hat{V}_{\rm ac}=-5.6$. The phase response is shown in the $x-y$ plane in Fig.~\ref{fig.phase_diagram}(a), where $x,y$ are the reduced densities such that $\rho_{\rm a} = (1-x)(1-y)\rho_{\rm t}$, $\rho_{\rm b} =  x(1-y) \rho_{\rm t}$ and $\rho_{\rm c} =  y \rho_{\rm t}$ (see Eq.~(\ref{eq::relative_densities})). Note that $x$ corresponds to the relative concentration of polymer b with respect to the total amount of polymers, while $(1-x)$ the relative concentration of polymer a. On the other hand, $y$ provides the amount of colloids relative to the total amount of system components (both polymer species and colloids). In panel (b) of Fig.~\ref{fig.phase_diagram} we show the coexisting colloidal densities as a function of $x$.  The coexisting phases are shown by upper and lower triangles, respectively, and are connected by tie lines (isobars). The color of the tie lines corresponds to the total density of the polymers, $\bar\rho_{\rm p}$, for which $\bar{\rho}_a= \bar{\rho}_b$ . The critical point, denoted by the full circle, is found at $\bar{\rho}_{\rm a}= \bar{\rho}_{b \rm }=0.1125$ and $\bar{\rho}_{\rm c} = 0.09$. This is slightly smaller than the value obtained by considering only the presence of polymers, which we find to be $\bar{\rho}_{\rm a}= \bar{\rho}_{b \rm }=0.135$. This clearly underlines the effect of the colloidal species in the phase separation of polymers, although being at a relative low concentration. In panel (b) we see that as the total polymer density $\bar\rho_{\rm p}$ is increased, the colloids partition more strongly within phase-I. This is expected, as the colloids favor the polymers making up said phase.

\begin{figure}[h!]
\includegraphics[width=\linewidth]{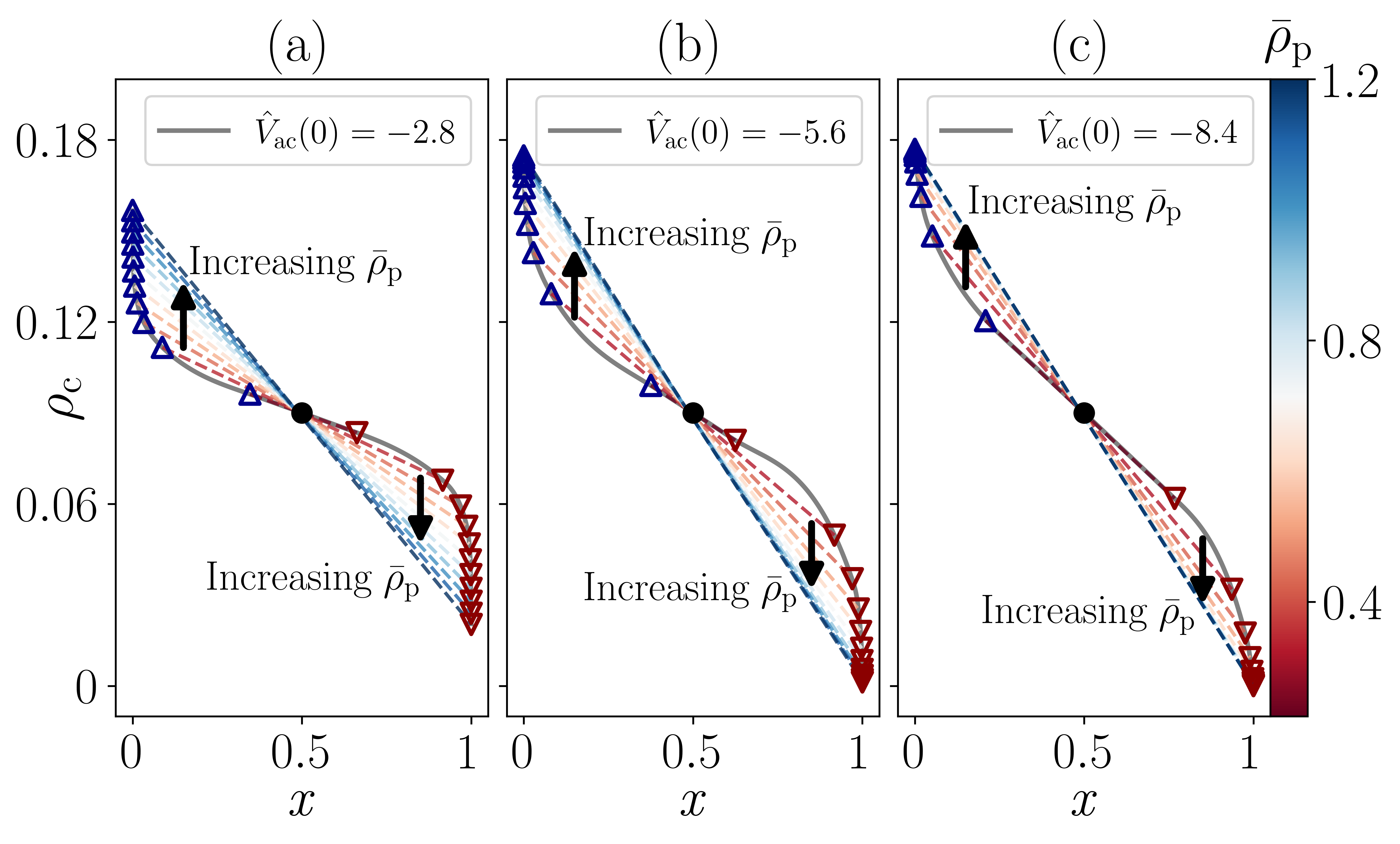}
  \caption{Partitioning of colloids driven for the same system as in Fig.~\ref{fig.phase_diagram} yet for (a) $\hat{V}_{\rm ac} = -2.8$, (b) $\hat{V}_{\rm ac} = -5.6$ and (c) $\hat{V}_{\rm ac}=-8.4$. The upper triangles indicate phase-I (polymer a-rich phase), while the lower triangles indicate phase-II (polymer b-rich phase). Coexisting points are connected by isobars, where their color corresponds to the value of $\bar\rho_{\rm p}$.}
\label{fig.4_4}
\end{figure}

For the same system configuration above,  we investigate the effect of the surface attraction strength between polymer a and colloids, $\hat{V}_{\rm ac}$. In Fig.~\ref{fig.4_4} we show the coexisting densities of the colloids in the two phases for $\hat{V}_{\rm ac}=-2.8$ (panel (a)), $\hat{V}_{\rm ac}=-5.6$ (panel (b)) and $\hat{V}_{\rm ac}=-8.4$ (panel (c)), respectively. As expected, as $\hat{V}_{\rm ac}$ is increased, the colloids partition more and more within phase-I. In line with the finding in Fig.~\ref{fig.phase_diagram}(b), this effect is more pronounced as the total polymer density is increased.  We also find a small deviation of the coexistence points from $x=0.5$ to smaller values of $x$ as $\hat{V}_{\rm ac}$ is increased, indicating a small enhancement in the polymer phase separation driven by the presence of the colloids. This is in line with previous findings in \cite{SCACCHI20251135}.  

\begin{figure}
\includegraphics[width=\linewidth]{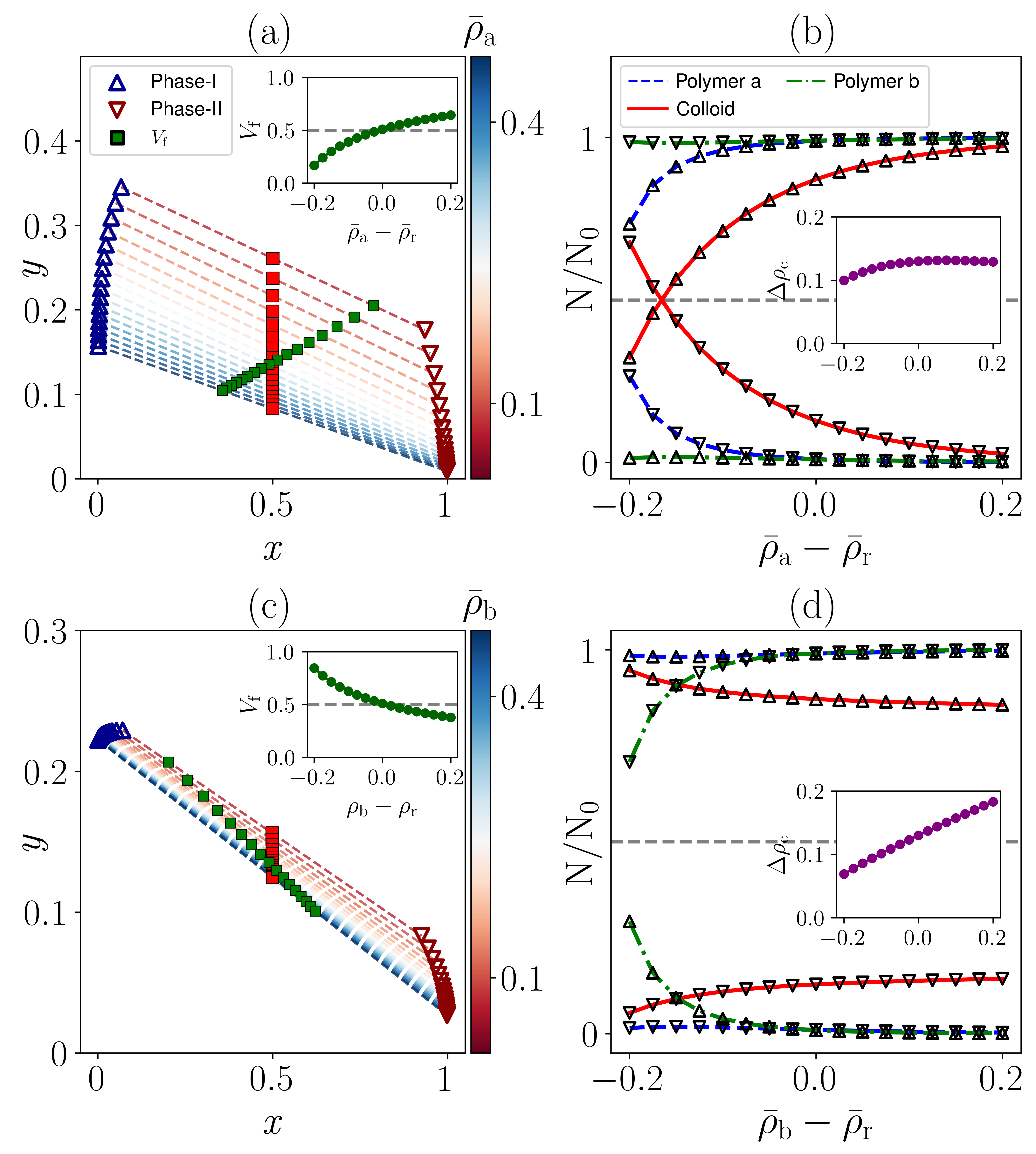}
\caption{System interacting as in Fig.~\ref{fig.phase_diagram}. Here $\rho_{\rm r} = 0.27$ and $\bar\rho_{\rm c} = 0.09$ are fixed. In (a), phase response, and in (b) partitioning of the system components for varying $\bar\rho_{\rm a}$. In (c), phase response, and in (d) partitioning of the system components for varying $\bar\rho_{\rm b}$. 
The upper triangles indicate phase-I (polymer a-rich phase), while the lower triangles indicate phase-II (polymer b-rich phase). Coexisting points are connected by isobars, where their color corresponds to the value of $\bar\rho_{\rm a}$ (panel (a)) and $\bar\rho_{\rm b}$ (panel (c)). The green squares represent the value of $V_{\rm f}$, i.e. the volume ratio of phase-I, while the red squares represent $V_{\rm f}=0.5$. The differences in the coexisting densities of the colloids, $\Delta \rho_{\rm c}$, are shown in the insets of panel (b) and (d). Dashed horizontal lines shown in panel (b) and (d) represent $N/N_{\rm 0} = 0.5 $. }
\label{fig.4_5}
\end{figure}

So far, we focused on the symmetric case $\bar{\rho}_{\rm a} = \bar{\rho}_{\rm b}$. In order to investigate the effect of the relative amount of the two polymer species, we now fix the colloids density  $\bar\rho_{\rm c} = 0.09$, and one polymer species in turns. The results are shown in Fig.~\ref{fig.4_5}, which illustrates a simple case of composition-driven phase separation, similar to those that can be observed in biological systems, such as, e.g., protein solutions~(see, e.g.,~\cite{farag2023phase}). Specifically, in panels (a) and (b), we fix $\bar\rho_{\rm b} = \rho_{\rm r} = 0.27$ and vary $\bar\rho_{\rm a}$ around the reference value, $\rho_{\rm r}$. As one can see in panel (a), the difference between the coexisting densities of the colloids in the two phases remain almost unaltered (the tielines are practically parallel). This can also be seen in the inset of panel (b)). On the other hand, their value changes, and correlates with the value of $\bar\rho_{\rm a}$. The values of the coexisting densities of the polymers, however, depends on $\bar\rho_{\rm a}$, as expected. To compensate the change in the density of species a, a shift in $V_{\rm f}$ away from value of the symmetric case, $V_f = 0.5$, is observed (see also inset of panel (a)). Figure \ref{fig.4_5}(b) shows the variation in the number of particles, normalized by their total amount in the system, $N_0$, of the different system components as a function of $\bar\rho_a$ in the two phases.  When $\bar\rho_{\rm a}$ is small, polymer a barely phase separate, yet partitioning is quickly increased as $\bar\rho_a$ is increased. Interestingly, there is a cross-over of the partitioning of the colloids. In fact, for small values of $\bar\rho_a$, colloids prefer phase-II. This trend is then reversed as $\bar\rho_a$ is increased. The partitioning of the colloids correlates with an increased volume of phase-I. Nevertheless, as already mention above, the coexisting densities of the colloids in the two phases vary only marginally (see inset of panel (b)). On the other hand, the relative amount of polymer b in the two phases is practically unaffected. 

Figure \ref{fig.4_5}(c) and (d) show the effect of varying $\bar\rho_{\rm b}$ while keeping $\bar\rho_{\rm a}=\rho_{\rm r} = 0.27$ and $\bar\rho_{\rm c}$ constant. Also here we see a strong dependency of the relative volume of the two phases (through the value of $V_{\rm f}$), as a function of $\bar\rho_{\rm b}$. As expected for these conditions, the volume of phase-II increases with $\bar\rho_{\rm b}$. This is shown in panel (c) and its inset. This, in turn, is expected to have an effect on the partitioning of the colloids. When considering the number of colloids in the different phases, the change seems marginal, as shown in panel (d). However, if one considers the coexisting densities (or the difference between them) of the colloids, shown in the inset of panel (d), the effect is relatively important. As in panels (a) and (b), we observe similar trends when it comes to the partitioning of the two polymer species.

\begin{figure}[h!]
\includegraphics[width=\linewidth]{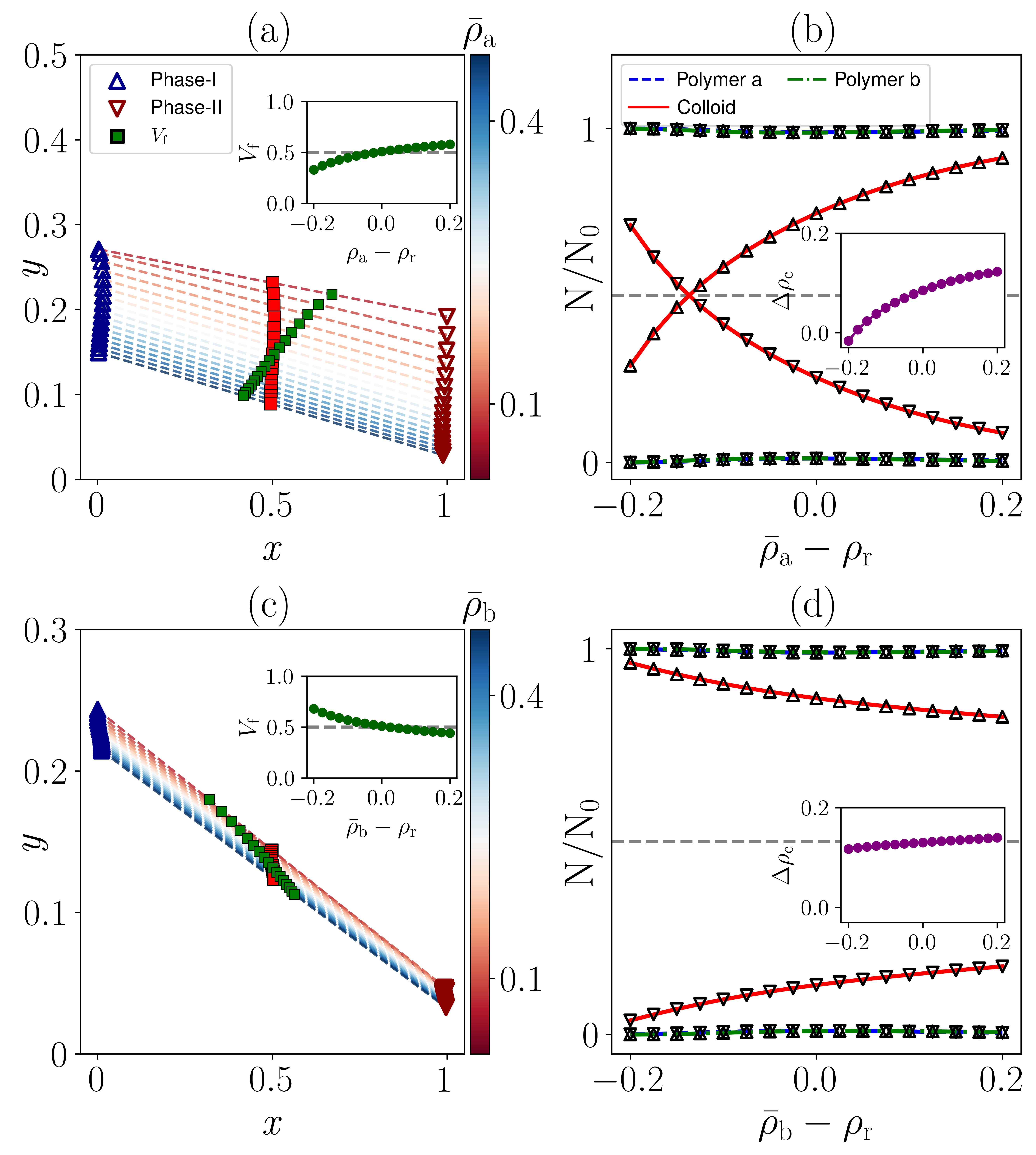}
  \caption{System interacting as in Fig.~\ref{fig.phase_diagram} with the exception of $R_{ij}$ following Eq.~(\ref{equ.meanpolymer_size}). Here $\rho_{\rm r} = 0.27$ and $\bar\rho_{\rm c} = 0.09$ are fixed. In (a), phase response, and in (b) partitioning of the system components for varying $\bar\rho_{\rm a}$ under the constraint $\varphi_{\rm a}= \frac{4}{3} \pi R_{\rm aa}^3 \bar\rho_{\rm a}=3.2$ and $\varphi_{\rm b}= \frac{4}{3} \pi R_{\rm bb}^3 \bar\rho_{\rm b}=3.2$ with varying $R_{\rm aa}$. 
  In (c), phase response, and in (d) partitioning of the system components for varying $\bar\rho_{\rm b}$ under the constraint $\varphi_{\rm a}= \frac{4}{3} \pi R_{\rm aa}^3 \bar\rho_{\rm a}=3.2$ and $\varphi_{\rm b}= \frac{4}{3} \pi R_{\rm bb}^3 \bar\rho_{\rm b}=3.2$ with varying $R_{\rm bb}$. Note that $\sigma_{\rm bb} = \sigma {R_{\rm bb}}/{R_{\rm r}}$, where $R_{\rm r}=\sqrt{2}\sigma$. 
The upper triangles indicate phase-I (polymer a-rich phase), while the lower triangles indicate phase-II (polymer b-rich phase). Coexisting points are connected by isobars, where their color corresponds to the value of $\bar\rho_{\rm a}$ (panel (a)) and $\bar\rho_{\rm b}$ (panel (c)). The green squares represent the value of $V_{\rm f}$, i.e. the volume ratio of phase-I, while the red squares represent $V_{\rm f}=0.5$. The differences in the coexisting densities of the colloids, $\Delta \rho_{\rm c}$, are shown in the insets of panel (b) and (d). Dashed horizontal lines shown in panel (b) and (d) represent $N/N_{\rm 0} = 0.5 $.
}
\label{fig.4_6}
\end{figure}
Next, we address the effect of polymer species densities by maintaining the occupied volume fractions fixed. The interactions between the system components are the same as above, with the exception of $R_{ij}$, which will change accordingly and following the sum rule in Eq.~(\ref{equ.meanpolymer_size}). Note that as the size of the polymers, $R_{ij}$, is changed, also the value of $\sigma_{ij}$ has to change accordingly. Specifically, $\sigma_{ii} = \sigma {R_{ii}}/{R_{\rm r}}$ for $i={\rm a, b}$, where $R_{\rm r}=\sqrt{2}\sigma$. 
This variation corresponds to a case in which the polymerization (or length of the polymers) is controlled. Firstly, we set $\varphi_{\rm a}= \frac{4}{3} \pi R_{\rm aa}^3 \bar\rho_{\rm a}=3.2$ and $\varphi_{\rm b}= \frac{4}{3} \pi R_{\rm bb}^3 \bar\rho_{\rm b}=3.2$, yet allowing both $R_{\rm aa}$ and $\bar\rho_{\rm a}$ to vary. Figure~\ref{fig.4_6}(a) shows the change in coexistence densities for different values of $\bar\rho_{\rm a}$ with respect to the reference density $\rho_{\rm r}=0.27$. The coexisting densities of the polymers do not show a strong dependency on $\bar\rho_{\rm a}$, yet the value of $V_{\rm f}$ increases importantly with the latter. The polymer partitioning also remains practically unchanged, as can be seen in panel (b). This is, however, not the case in Fig. \ref{fig.4_5}(b), where the partitioning of polymer a was strongly affected by its density. On the other hand, the distribution of colloids is strongly affected by the increase of $\bar\rho_{\rm a}$ (which corresponds to a decrease in $\sigma_{\rm aa}$). This is true both in terms of particle distribution (panel (a)), as well as in terms of coexisting densities (inset of panel (b)).
Secondly, we set $\varphi_{\rm a}= \frac{4}{3} \pi R_{\rm aa}^3 \bar\rho_{\rm a}=3.2$ and $\varphi_{\rm b}= \frac{4}{3} \pi R_{\rm bb}^3 \bar\rho_{\rm b}=3.2$, yet allowing both $R_{\rm bb}$ and $\bar\rho_{\rm b}$ to vary. Also here, the effect on polymer coexistence is marginal, however, the change of corresponding volume of the two phases is important (see Fig. \ref{fig.4_6}(c)). Similar to panel (b), also the partitioning of the polymers is marginally affected by the value of $\bar\rho_{\rm b}$. This is again in contrast to Fig.~\ref{fig.4_5}(d), where the partitioning of polymer b is increasing as  $\bar\rho_{\rm b}$ is increased. Interestingly, as $\bar\rho_{\rm b}$ is increased, the partitioning of the colloids is weakened (see panel (d)), while $\Delta\rho_{\rm c}$ is increased.

\begin{figure}
\includegraphics[width=\linewidth]{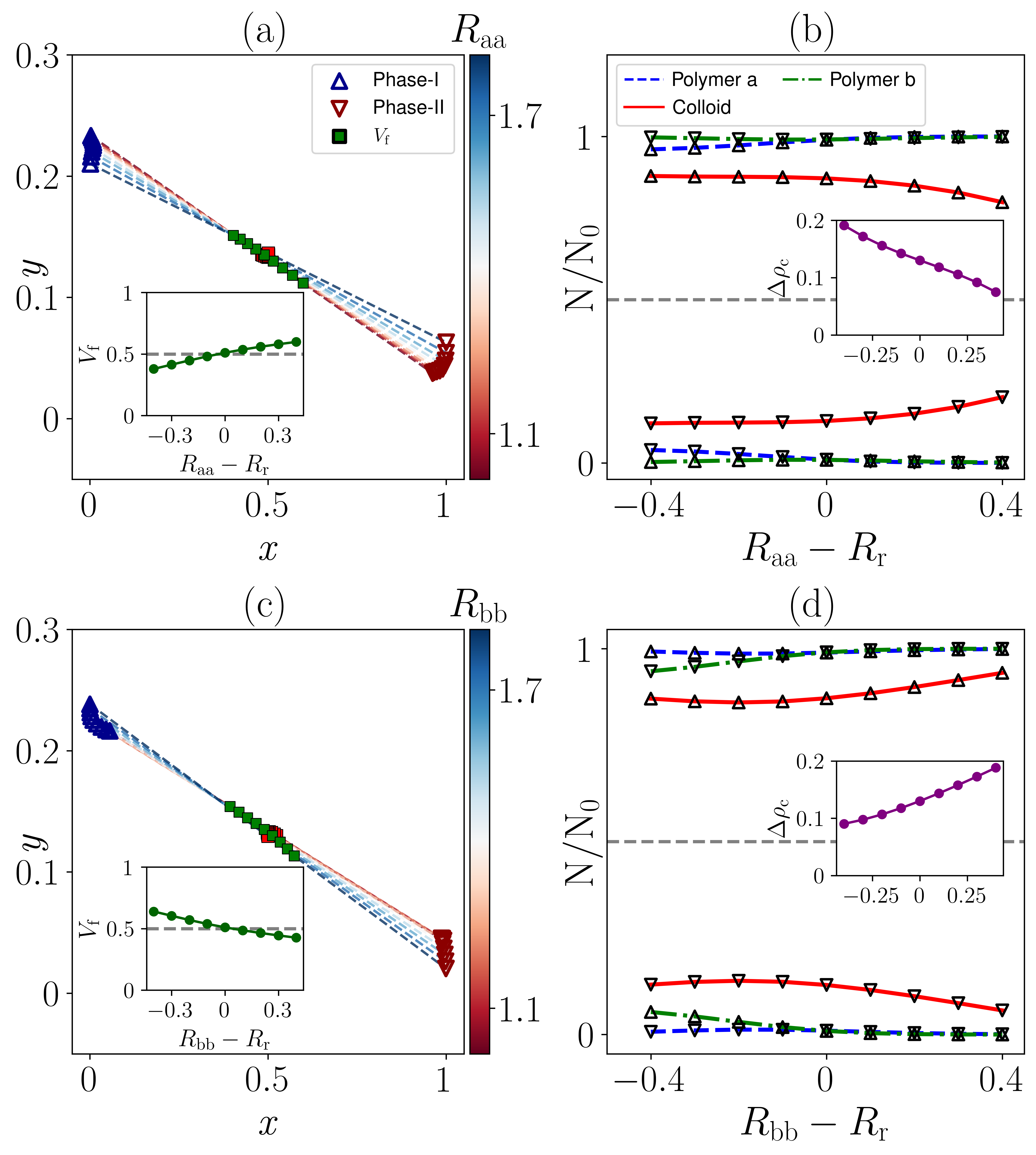}
  \caption{Here $\bar{\rho}_{\rm a} = \bar{\rho}_{\rm b} = 0.27$ and $\bar{\rho}_{\rm c} = 0.09$ with $ \hat{V}_{\rm ac} = -5.6$.
Coexistence densities (a) and partitioning (b) of system components for varying sizes of polymer a, where $R_{\rm bb}=\sqrt{2}\sigma=R_{\rm r}$. Note that $R_{\rm ab}$ follows from Eq.~(\ref{equ.meanpolymer_size}). Coexistence densities (c) and partitioning (d) of system components for varying sizes of polymer b, where $R_{\rm aa}=\sqrt{2}\sigma=R_{\rm r}$. The upper triangles indicate phase-I (polymer a-rich phase), while the lower triangles indicate phase-II (polymer b-rich phase). Coexisting points are connected by isobars, where their color corresponds to the value of $R_{\rm aa}$ and $R_{\rm bb}$ in panel (a) and (c), respectively. The green squares represent the value of $V_{\rm f}$, i.e. the volume ratio of phase-I, while  the red squares represent $V_{\rm f}=0.5$. Full circle represents the critical point. Dashed horizontal lines shown in panel (b) and (d) represent $N/N_{\rm 0} = 0.5 $. }
\label{fig.4_7}
\end{figure}
Finally, in order to explore the effect of asymmetry in polymer sizes, corresponding to asymmetry in depletion forces, we change the size of one polymer species while fixing the size of the other. Such variation is relevant, as it has been recently shown experimentally that system containing mostly short polymers exhibit a narrower two-phase region, together with a reduced density differences~\cite{kamo2025impact}. Also here, we keep the average densities fixed as before ($\bar\rho_{\rm a} = \bar\rho_{\rm b} = 0.27$ and $\bar\rho_{\rm c} = 0.09$) and the interaction strength between colloids and polymer a at $\hat{V}_{\rm ac} = -5.6$. With $R_{\rm r}=\sqrt{2}\sigma$ being the reference polymer size, in Fig.~\ref{fig.4_7}(a),(b) we investigate the effect of $R_{\rm aa}$ on phase separation and partitioning, respectively, while keeping $R_{\rm bb}=R_{\rm r}$. Note that any change in polymer size implies a change in the value of $\sigma_{ii}$, which is linearly rescaled correspondingly, i.e. $\sigma_{ii} =  \sigma  
R_{ii}/R_{\rm r}$, where $i=\rm a,b$. In panel (a) we see that the coexisting densities of the polymers are practically unaffected by the value of $R_{\rm aa}$, while the phase boundary, described by $V_{\rm f}$, shifts. Specifically, when $R_{\rm aa}>R_{\rm r}$, phase-I (polymer a-rich phase) increases in volume. Nevertheless, the partitioning of each species (shown in panel (b)), remain practically unaffected, with a slight variation for small values of $R_{\rm aa}$ in $N/N_0$ for polymer a. However, this small change points towards a narrowing of the phase separation region of the polymer mixture, in line with Ref.~\citenum{kamo2025impact}. On the other hand, the difference in coexisting densities of the colloids decreases importantly as $R_{\rm aa}$ is increased. 
Similarly, in Fig.~\ref{fig.4_7}(c),(d) we study the same response yet as a function of $R_{\rm bb}$. Also in this case, as shown in panel (c), the values of the coexisting densities are barely affected, yet the phase boundary shifts such that the volume of phase-II increases as $R_{\rm bb}$ is increased. Also similar to the previous case, the partitioning of the different system components are practically unchanged (panel (d)). On the other, the inset of panel (d) shows that the difference of coexisting densities of the colloids correlates with the size of polymer b.

\subsubsection*{Effect of colloids properties on the phase response}

\begin{figure}[h!]
\includegraphics[width=\linewidth]{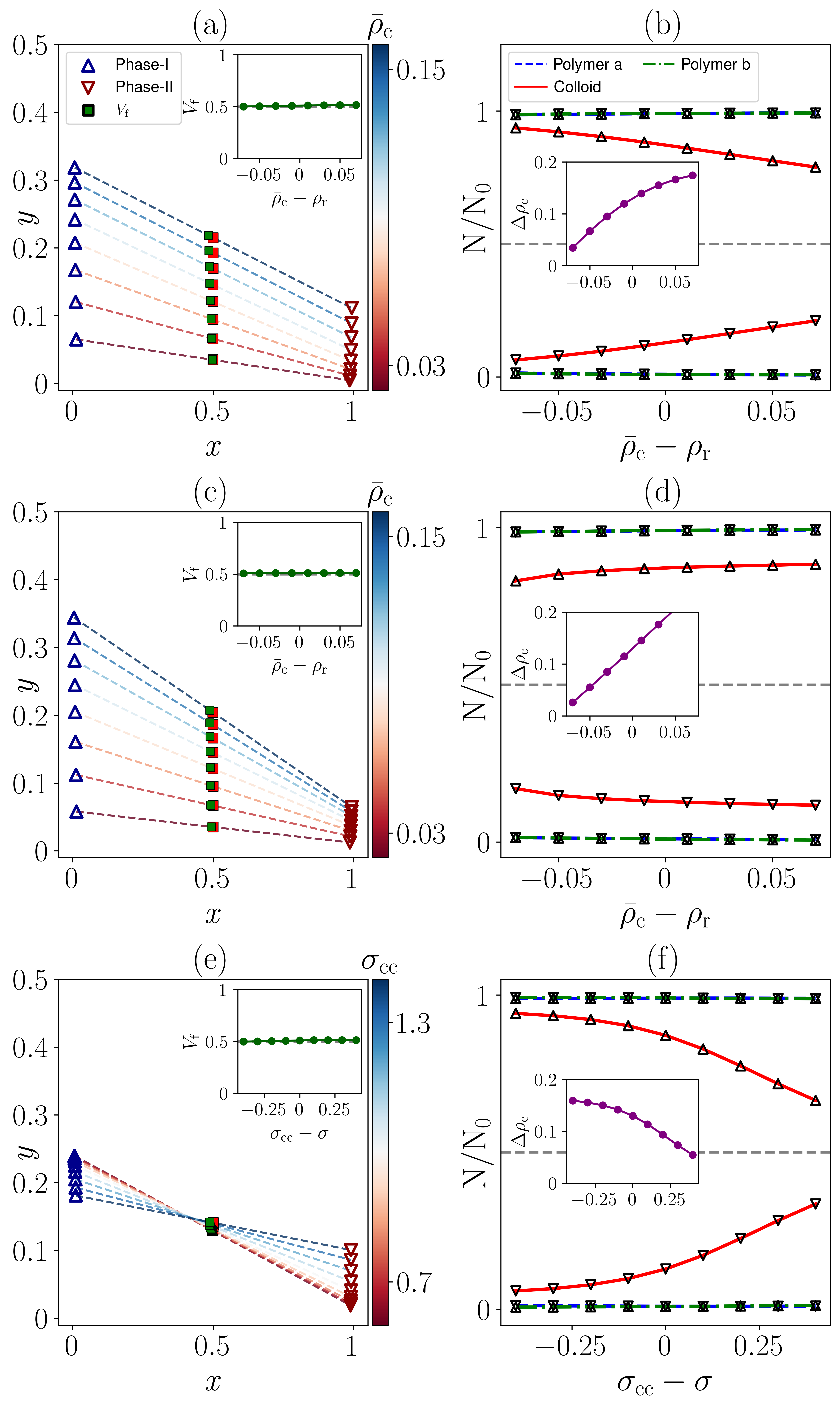}
  \caption{Here $R_{\rm aa} = R_{\rm bb} = R_{\rm ab} =  \sqrt{2}\sigma$, $\epsilon_{\rm aa} = \epsilon_{\rm bb} = 2.0, \epsilon_{\rm ab} =  2.5$, $\sigma_{\rm aa} = \sigma_{\rm bb} = \sigma$ and $\hat{V}_{\rm ac}=-5.6$. Additional constraints include $\bar\rho_{\rm a} = \bar\rho_{\rm b} = 0.27$, and reference density $\rho_{\rm r}=0.09$.
  In (a), phase response, and in (b) partitioning of the system components for varying $\bar\rho_{\rm c}$. In (c), phase response, and in (d) partitioning of the system components for varying $\bar\rho_{\rm c}$ with fix volume fraction $\varphi_{\rm c}= \frac{4}{3} \pi (\sigma_{\rm cc}/2)^3 \bar\rho_{\rm c}= 0.04712$, with $\bar \rho_{\rm c}=0.09$ ($\sigma_{\rm cc}$ varies accordingly).
  In (e), phase response, and in (f) partitioning of the different system components, for $\bar\rho_{\rm c}=0.09$ for varying colloids size, $\sigma_{\rm cc}$. The upper triangles indicate phase-I (polymer a-rich phase), while the lower triangles indicate phase-II (polymer b-rich phase). Coexisting points are connected by isobars (see colorbar). The green squares represent the value of $V_{\rm f}$, while the red squares represent $V_{\rm f}=0.5$. The differences in the coexisting densities of the colloids, $\Delta \rho_{\rm c}$ shown in the insets of panel (b), (d) and (f). Dashed horizontal lines shown in panel (b), (d) and (f) represent $N/N_{\rm 0} = 0.5 $.
}
\label{fig.colloids phase response}
\end{figure}

To examine the effect of physical properties of the colloidal species on the system phase response and partitioning, we investigate the effects of colloid density and relative sizes. Beside the variations specified hereafter, the interaction describing the system are: $R_{\rm aa} = R_{\rm bb} = R_{\rm ab} =  \sqrt{2}\sigma$, $\epsilon_{\rm aa} = \epsilon_{\rm bb} = 2.0, \epsilon_{\rm ab} =  2.5$, $\sigma_{\rm aa} = \sigma_{\rm bb} = \sigma$ and $ \hat{V}_{\rm ac}=-5.6$. For all cases addressed here, the polymer densities are fixed to $\bar\rho_{\rm a} = \bar\rho_{\rm b} = 0.27$. Here we define the reference colloidal density, $\rho_{\rm r}=0.09$. Specifically, in panels (a) and (b) of Fig.~\ref{fig.colloids phase response} we investigate the effect of the colloidal density, $\bar\rho_{\rm c}$, on the phase response and partitioning, respectively, of colloids with size $\sigma_{\rm cc}=\sigma_{\rm aa}=\sigma_{\rm bb}=\sigma$. From both (a) and (b) we can see that the amount of colloids has a marginal effect on the coexisting densities of the polymers (almost no variation in the value of $x$, and the change in $y$ is compensated by the same change in $\rho_{\rm t}$), as well as a marginal influence on $V_{\rm f}$. However, the difference in coexisting densities of the colloids clearly increases as shown in the inset of panel (b). On the other hand, partitioning of the colloids between the two phases decreases with increasing $\bar\rho_{\rm c}$, as the finite volume of polymers limits their ability to attract colloids in the same proportion (red curve in panel (b)). In Fig.~\ref{fig.colloids phase response}(c),(d) we address the effect of colloidal density, under the constraint of constant occupied volume fraction, which is achieved by changing $\sigma_{\rm cc}$ accordingly, as the volume fraction of the colloids is given by $\varphi_{\rm c}= \frac{4}{3} \pi (\sigma_{\rm cc}/2)^3 \bar\rho_{\rm c}$. Here, $\varphi_{\rm c}=0.04712$. We find that changing the colloid density while keeping the packing fraction constant does not significantly alter the polymer phase behavior (panel (c)). However, colloid partitioning increases when a larger number of smaller colloids are present (higher values of $\bar\rho_{\rm c}$), as shown in panel (d)). This is opposite to the trend in Fig.~\ref{fig.colloids phase response}(b), highlighting the role of colloid size in partitioning, despite both polymer species exerting the same depletion force ($\sigma_{\rm aa} = \sigma_{\rm bb}$). Finally, in panels (e) and (f), for fixed $\bar\rho_{\rm c}$, we vary the size of the colloids, $\sigma_{\rm cc}$, around the reference value, $\sigma$. Also here, as can be seen in both panel (c) and (d), the effect on polymer coexistence is marginal (same for $V_{\rm f}$). On the other hand, colloid size has a strong impact on colloid partitioning itself. As the colloid size decreases, partitioning increases (panel (f)), even if the colloids exert the same depletion force on both the polymers ($\sigma_{\rm aa} = \sigma_{\rm bb}$). This is due to a shift in the competition between depletion forces and attractive forces, which makes attractive forces more dominant for smaller colloids. This last results underlines the importance of polymer-colloid size ratio in controlling partitioning.

\subsection{Inhomogeneous properties }
We conclude by examining the inhomogeneous density profiles of the system for a specific case. In addition to providing a typical response, we also compare our cDFT results with particle-based simulations. To this end, we confine the mixture between two parallel walls perpendicular to the $z$-axis, described by the potential
\begin{equation}
u_{\rm w}(z) =  \epsilon \left( \frac{2}{15}\left(\frac{\sigma}{z}\right)^{9} - \left(\frac{\sigma}{z}\right)^{3}  \right),
\label{eq.4_35}
\end{equation}
where $z$ is the distance from the wall. We start by defining the interactions and the thermodynamic state point for particle-based simulations. Here the polymer-polymer interactions are modeled using $R_{\rm aa} = R_{\rm bb} = R_{\rm ab} =  \sqrt{2}\sigma$ and $\epsilon_{\rm aa} = \epsilon_{\rm bb} = 2.0, \epsilon_{\rm ab} =  2.5$. The colloid-colloid and colloid-polymer interactions follow Eq.~(\ref{eq::coll-pol_simulation}), with the parameters specified in Table~\ref{tab:interaction_inhom}. 
\begin{table}[h]
\begin{tabular}{|c|c|c|c|c|c|}
\hline
\multicolumn{6}{|c|}{Colloid-Colloid and Colloid–Polymer} \\ \hline
Species & u & v & $\epsilon_{ij}$ & $\sigma_{ij}$ & $r_{\rm cutoff}$ \\ \hline
cc       & 12 & 6  & $1$      & $1$ & $2^{1/6}$ \\
bc       & 12 & 6  & $1$      & $1$ & $2^{1/6}$ \\
ac       & 36 & 18 & $1.6358$ & $1$ & $5$ \\ \hline
\end{tabular}
\caption{Colloid-colloid and colloid–polymer interaction parameters for the inhomogeneous system in particle-based simulations.}
\label{tab:interaction_inhom}
\end{table}
\begin{figure}[h]
\includegraphics[width=\linewidth]{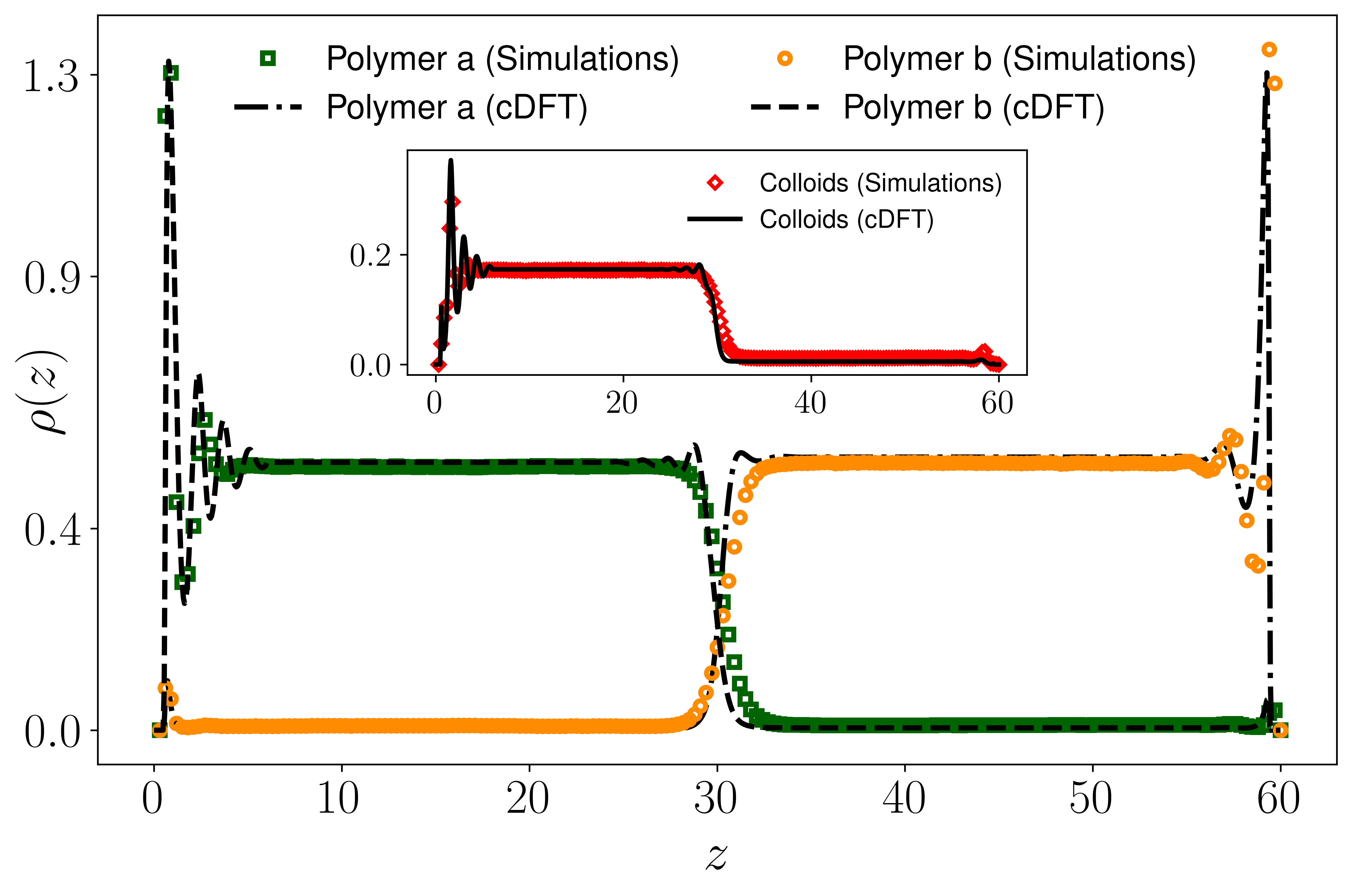}
\caption{Density profiles for a ternary colloid–polymer mixture.  Polymer-polymer interactions are modeled using $R_{\rm aa} = R_{\rm bb} = R_{\rm ab} =  \sqrt{2}\sigma$ and $\epsilon_{\rm aa} = \epsilon_{\rm bb} = 2.0, \epsilon_{\rm ab} =  2.5$. Colloid-colloid and colloid-polymer interactions follow Eq.~(\ref{eq::coll-pol_simulation}), with the parameters specified in Table~\ref{tab:interaction_inhom}. Average densities are $\bar\rho_{\rm a}=0.27$, $\bar\rho_{\rm b}=0.27$, and $\bar\rho_{\rm c}=0.09$. Confinement is imposed with Eq.~(\ref{eq.4_35}). Solid lines denote cDFT predictions, while dashed lines correspond to particle-based Brownian dynamics simulations (as in~\cite{SCACCHI20251135}).
}
\label{fig.4_8}
\end{figure}
The simulations are run in a box with size $15\sigma \times15\sigma\times60\sigma$ for a total of $2\times10^8$ steps, of which the first half is considered as equilibration time. Also here, $d\tau=10^{-4}\sigma^2D^{-1}$, where $D$ is the particle self-diffusion constant.

Also here, the colloid has an affinity with polymer a, while colloid-colloid and colloid-polymer b interactions are purely repulsive. We focus on the following densities: $\bar\rho_{\rm a}=0.27$, $\bar\rho_{\rm b}=0.27$, and $\bar\rho_{\rm c}=0.09$. The particle-based simulations results are obtained by means of standard Brownian dynamics simulations, as in~\cite{SCACCHI20251135}. The simulations are run in a box with size $15\sigma \times15\sigma\times60\sigma$ for a total of $2\times10^8$ steps, of which the first half is considered as equilibration time. Also here, the dimensionless time step is set to $d\tau=10^{-4}\sigma^2D^{-1}$, where $D$ is the particle self-diffusion constant. Equilibration is assessed by observing the individual density profiles until they show no further changes.

Finding the corresponding cDFT profiles requires two main steps. In the first step, one has to solve for bulk coexistence by specifying the average densities, $\bar\rho_{\rm a}$, $\bar\rho_{\rm b}$, and $\bar\rho_{\rm c}$. The polymers can be mapped directly from the simulations, while colloid-colloid and colloid-polymer interactions require the calculation of effective sizes, which is done using Eq.~(\ref{eq::bh_radius}). The corresponding BH diameter for this specific system are: $\sigma^{\rm BH}_{\rm cc}=\sigma^{\rm BH}_{\rm bc}=0.973$ and $\sigma^{\rm BH}_{\rm ac}=0.986$. On the other hand, as already mentioned in the case of the comparison between bulk cDFT and particle-based simulations, the contribution from the attractive interactions within the MF approximation is not uniquely defined. To overcome this issue, we vary the value of $\hat{V}_{\rm ac}$ in cDFT to match the coexisting colloidal densities~\cite{velasco1989wetting}. For the case studied here, we find the best fit for $\hat{V}_{\rm ac}=-9.132$. We can now carry out the second step, which consists in solving the inhomogeneous cDFT with the initial guesses obtained for bulk \cite{bruno1987wetting}. This is done by solving Eq.~(\ref{eq::inhomog_density}) iteratively, with the chemical potentials obtained from the bulk solution. The evaluation of $\mathcal{F}_{\rm exc}[\{ \rho_i(z)\}]$ requires integration over the $x$ and $y$ directions for the free-energy functionals described in Sec.~II C. 

We use a Picard iteration, namely
\begin{equation}
\rho_i^k(z) = \alpha \rho^{\rm new}_i(z) + (1-\alpha) \rho_i^{k-1}(z),
\end{equation}
where $k$ is the iteration step and $\alpha$ is a mixing parameter. We employ $\alpha = 0.001$ for $\bar\rho_c < 0.12$ and $\alpha < 0.00001$ for $\bar\rho_c \geq 0.12$. Further details, including the numerical implementation, are provided in the SI.

The density profiles for this specific case are shown in Fig.~\ref{fig.4_8}. We find a satisfactory matching between the particle-based simulations and our cDFT. Most satisfactory is the matching of the coexisting densities, including both polymer and colloidal species, together with the first density peaks at the walls. On the other hand, the position and shape of the interface between the two phases is captures only qualitatively. Specifically, the interface is sharper in cDFT, and shows additional peaks which are inexistent in the particle-based simulations, as previously observed~\cite{PhysRevE.103.042103}. Nevertheless, although the system is highly complex and the simulations rely on soft potentials while cDFT assumes hard interactions, the theory provides valuable physical insight.

%Now, to investigate further on the effect of colloids' partitioning, on the interfacial properties, we computed surface tension given as,

%\begin{equation}
%\gamma = \int_{-\infty}^{\infty} (P_{zz}(z) - \frac{1}{2}[P_{xx}(z) +P_{yy}(z)])  \,dz
%\end{equation}

%Fig. \ref{fig.4_9} shows variation in $\gamma$ with increasing $\rho_t^c$ in the system. The surface tension increases with the increase in the total number of colloids in the system and then stagnate around a constant value. It follows the trend as shown by the $\Delta \rho_c$, indicating about the same being the primary factor, affecting the interfacial properties rather than $\rho_t^c$. In the inset, we have shown $\Delta \omega = \int_{-\infty}^{\infty} (\omega(z) - \omega_b) dz $. Where $\omega = \sigma^3\Omega/V$ is grand potential per unit volume. Fluctuation of $\Delta \omega$ close to the interface on the side of colloid rich phase-Increases (inset in Fig. \ref{fig.4_9}) with increasing the partitioning due to more and more accumulation of colloids (red curve in the Fig. \ref{fig.4_8}). In other words system with the high degree of partitioning in colloids shows more stability close to the interface.

%\begin{figure}
%\includegraphics[height=6cm,width=9cm]{surface_tension_with_inset.png}
%  \caption{Interface properties.}
%\label{fig.4_9}
%\end{figure}

\section{Conclusions and perspective}

In this work, we proposed a general scheme for studying the bulk multiphase coexistence for a mixture composed by n polymer and m colloidal species with different affinities. This scheme can be applied to bimolecular condensates~\cite{holehouse2025molecular}, where several polymeric fluid phases coexist with other dispersed macromolecules. In the scheme, we proposed an enhanced mean field approach within the classical density functional theory formalism~\cite{Evans01041979}. For the case of two demixing polymer species and one dispersed colloidal component affine to only one polymer type, our scheme has been benchmarked with particle-based simulations. We found both qualitative and quantitative results (see Figs.~\ref{fig.phase_response} and \ref{fig.4_8}). This specific case is relevant in the context of water purification by means of aqueous two-phase systems~\cite{iqbal2016aqueous, ASSIS2021121697}. We found that the partitioning of the colloidal species is highly sensitive to the strength of affinity towards the affine polymer species, as well as its density (the higher the density, the stronger the partitioning), together with the size ratio between polymers and colloids (smaller colloids partition more). On the other hand, we found marginal influence on the partitioning of colloids coming from variations in the density of the non-affine polymer species, and from the asymmetry in the two polymer species sizes. 

Notably, our findings resonate with previous studies on colloidal systems in binary solvent mixtures near their critical point~\cite{edison2015critical, edison2015phase}, which also reported rich phase behavior and highlighted the role of parameters such as colloid size, colloid wettability and {\it distance} from the critical point, in controlling phase behavior. Although the physical systems are quite different (binary solvents in their case and immiscible polymers in water in ours), the parallels highlight the broader relevance of these mechanisms across several soft matter systems. Another interesting case that shares similarities with our work is the recent study of ouzo~\cite{D4SM00332B}, which consists of a ternary mixture of water, alcohol and trans-anethole oil. In this work, the experimental bulk phase diagram and interfacial properties have been benchmarked against a lattice DFT approach.

From an experimental perspective, in the case of soluble yet demixing polymers, it is safe to assume $\epsilon_{\rm aa}=\epsilon_{\rm bb}=2k_{\rm B}T$~\cite{10.1063/1.1344606}, while the value of $\epsilon_{\rm ab}$ should be fitted from experiments at different polymer concentrations. On the other hand, the affinity between the polymer species and the colloids can be estimated through experiments assessing the second virial coefficients, such as static light scattering \cite{ma2015determination}.

It is important to note that the current formalism is valid only for additive mixtures of spherical, charge-neutral constituents. Nevertheless, density functional theory approaches that address this gap have already been developed (see, e.g.,~\cite{schmidt2004rosenfeld, hartel2015fundamental, wittmann2016fundamental}). We hope that future work by the community will integrate such advances into our formalism, as extending our scheme to include these features, while challenging, remains feasible.

As a final note, to the best of our knowledge, this is the first time in which the affinity between a polymer species and colloids has been implemented within classical density functional theory.

%A general scheme for the calculation of the coexisting densities for a multiphase system presented along with the methods to solve it efficiently in a canonical ensemble. The effective volume correction in the mean-field leads to the correct phase response of the system with varying parameters. However, a deeper investigation in the validity of approach is still required, despite its success to produce the accurate result. cDFT system indicates that the increase in densities in a species rich phase does not indicate about the better partitioning, where the volume of the phase plays also an important role. The presented work is aimed to be used as a tool in the multi-component system. The application of FMT approach provides the scheme a large room for the customization of the same for the purpose of the study of anisotropic particles as well as non-additive colloidal mixture. The same scheme can be implemented to the study of purely colloidal system with attractive term treated in mean field and the repulsive term using FMT approach.  

%\section{Future perspective}
%\textcolor{blue}{I would like to write here what can be done for improving the approach. For example, better way for calculating the MF attraction (eventually connecting to B2). Also, what should be changed for anisotropic particles (both HC and FV term and attractive part of MF). If also polymers are anisotropic, than also MF for polymers is affected. And others, if you know others.}

%\textbf{Chemistry day for me is optional.}

%\vspace{5mm}
\section*{Acknowledgments}
This work was supported by the Jane and Aatos Erkko Foundation under the grant No. 230052 (A.S.). Computational resources by CSC IT Center for Finland are also gratefully acknowledged. The work was conducted within the \#SUSMAT profiling measure at the University of Turku, Finland. We are also thankful to Prof. Andy Archer for fruitful discussions.

\section*{Supplementary information}
The supplementary information, including the algorithms developed here, will be available after peer-review.

\bibliographystyle{apsrev4-1}
\bibliography{dft_new}

\begin{thebibliography}{62}%
\makeatletter
\providecommand \@ifxundefined [1]{%
 \@ifx{#1\undefined}
}%
\providecommand \@ifnum [1]{%
 \ifnum #1\expandafter \@firstoftwo
 \else \expandafter \@secondoftwo
 \fi
}%
\providecommand \@ifx [1]{%
 \ifx #1\expandafter \@firstoftwo
 \else \expandafter \@secondoftwo
 \fi
}%
\providecommand \natexlab [1]{#1}%
\providecommand \enquote  [1]{``#1''}%
\providecommand \bibnamefont  [1]{#1}%
\providecommand \bibfnamefont [1]{#1}%
\providecommand \citenamefont [1]{#1}%
\providecommand \href@noop [0]{\@secondoftwo}%
\providecommand \href [0]{\begingroup \@sanitize@url \@href}%
\providecommand \@href[1]{\@@startlink{#1}\@@href}%
\providecommand \@@href[1]{\endgroup#1\@@endlink}%
\providecommand \@sanitize@url [0]{\catcode `\\12\catcode `\$12\catcode `\&12\catcode `\#12\catcode `\^12\catcode `\_12\catcode `\%12\relax}%
\providecommand \@@startlink[1]{}%
\providecommand \@@endlink[0]{}%
\providecommand \url  [0]{\begingroup\@sanitize@url \@url }%
\providecommand \@url [1]{\endgroup\@href {#1}{\urlprefix }}%
\providecommand \urlprefix  [0]{URL }%
\providecommand \Eprint [0]{\href }%
\providecommand \doibase [0]{http://dx.doi.org/}%
\providecommand \selectlanguage [0]{\@gobble}%
\providecommand \bibinfo  [0]{\@secondoftwo}%
\providecommand \bibfield  [0]{\@secondoftwo}%
\providecommand \translation [1]{[#1]}%
\providecommand \BibitemOpen [0]{}%
\providecommand \bibitemStop [0]{}%
\providecommand \bibitemNoStop [0]{.\EOS\space}%
\providecommand \EOS [0]{\spacefactor3000\relax}%
\providecommand \BibitemShut  [1]{\csname bibitem#1\endcsname}%
\let\auto@bib@innerbib\@empty
%</preamble>
\bibitem [{\citenamefont {Hyman}\ \emph {et~al.}(2014)\citenamefont {Hyman}, \citenamefont {Weber},\ and\ \citenamefont {Jülicher}}]{cell_seperation}%
  \BibitemOpen
  \bibfield  {author} {\bibinfo {author} {\bibfnamefont {A.~A.}\ \bibnamefont {Hyman}}, \bibinfo {author} {\bibfnamefont {C.~A.}\ \bibnamefont {Weber}}, \ and\ \bibinfo {author} {\bibfnamefont {F.}~\bibnamefont {Jülicher}},\ }\href {\doibase https://doi.org/10.1146/annurev-cellbio-100913-013325} {\bibfield  {journal} {\bibinfo  {journal} {Annual Review of Cell and Developmental Biology}\ }\textbf {\bibinfo {volume} {30}},\ \bibinfo {pages} {39} (\bibinfo {year} {2014})}\BibitemShut {NoStop}%
\bibitem [{\citenamefont {Balakrishnan}\ and\ \citenamefont {Kenworthy}(2024)}]{doi:10.1021/jacs.3c10132}%
  \BibitemOpen
  \bibfield  {author} {\bibinfo {author} {\bibfnamefont {M.}~\bibnamefont {Balakrishnan}}\ and\ \bibinfo {author} {\bibfnamefont {A.~K.}\ \bibnamefont {Kenworthy}},\ }\href {\doibase 10.1021/jacs.3c10132} {\bibfield  {journal} {\bibinfo  {journal} {Journal of the American Chemical Society}\ }\textbf {\bibinfo {volume} {146}},\ \bibinfo {pages} {1374} (\bibinfo {year} {2024})}\BibitemShut {NoStop}%
\bibitem [{\citenamefont {Mej{\'\i}as}\ \emph {et~al.}(2025)\citenamefont {Mej{\'\i}as}, \citenamefont {Seballos},\ and\ \citenamefont {Lafon-Hughes}}]{mejias2025liquid}%
  \BibitemOpen
  \bibfield  {author} {\bibinfo {author} {\bibfnamefont {D.}~\bibnamefont {Mej{\'\i}as}}, \bibinfo {author} {\bibfnamefont {V.}~\bibnamefont {Seballos}}, \ and\ \bibinfo {author} {\bibfnamefont {L.}~\bibnamefont {Lafon-Hughes}},\ }\href@noop {} {\bibfield  {journal} {\bibinfo  {journal} {Biophysical Reviews}\ ,\ \bibinfo {pages} {1}} (\bibinfo {year} {2025})}\BibitemShut {NoStop}%
\bibitem [{\citenamefont {Shin}\ and\ \citenamefont {Brangwynne}(2017)}]{doi:10.1126/science.aaf4382}%
  \BibitemOpen
  \bibfield  {author} {\bibinfo {author} {\bibfnamefont {Y.}~\bibnamefont {Shin}}\ and\ \bibinfo {author} {\bibfnamefont {C.~P.}\ \bibnamefont {Brangwynne}},\ }\href {\doibase 10.1126/science.aaf4382} {\bibfield  {journal} {\bibinfo  {journal} {Science}\ }\textbf {\bibinfo {volume} {357}},\ \bibinfo {pages} {eaaf4382} (\bibinfo {year} {2017})}\BibitemShut {NoStop}%
\bibitem [{\citenamefont {Georgiou}\ \emph {et~al.}(2025)\citenamefont {Georgiou}, \citenamefont {Hoogenboom},\ and\ \citenamefont {Kikuchi}}]{D4PY90151G}%
  \BibitemOpen
  \bibfield  {author} {\bibinfo {author} {\bibfnamefont {T.~K.}\ \bibnamefont {Georgiou}}, \bibinfo {author} {\bibfnamefont {R.}~\bibnamefont {Hoogenboom}}, \ and\ \bibinfo {author} {\bibfnamefont {A.}~\bibnamefont {Kikuchi}},\ }\href {\doibase 10.1039/D4PY90151G} {\bibfield  {journal} {\bibinfo  {journal} {Polymer Chemistry}\ }\textbf {\bibinfo {volume} {16}},\ \bibinfo {pages} {233} (\bibinfo {year} {2025})}\BibitemShut {NoStop}%
\bibitem [{\citenamefont {Rothschild}\ \emph {et~al.}(2024)\citenamefont {Rothschild}, \citenamefont {Averesch}, \citenamefont {Strychalski}, \citenamefont {Moser}, \citenamefont {Glass}, \citenamefont {Cruz~Perez}, \citenamefont {Yekinni}, \citenamefont {Rothschild-Mancinelli}, \citenamefont {Roberts~Kingman}, \citenamefont {Wu}, \citenamefont {Waeterschoot}, \citenamefont {Ioannou}, \citenamefont {Jewett}, \citenamefont {Liu}, \citenamefont {Noireaux}, \citenamefont {Sorenson},\ and\ \citenamefont {Adamala}}]{doi:10.1021/acssynbio.3c00724}%
  \BibitemOpen
  \bibfield  {author} {\bibinfo {author} {\bibfnamefont {L.~J.}\ \bibnamefont {Rothschild}}, \bibinfo {author} {\bibfnamefont {N.~J.~H.}\ \bibnamefont {Averesch}}, \bibinfo {author} {\bibfnamefont {E.~A.}\ \bibnamefont {Strychalski}}, \bibinfo {author} {\bibfnamefont {F.}~\bibnamefont {Moser}}, \bibinfo {author} {\bibfnamefont {J.~I.}\ \bibnamefont {Glass}}, \bibinfo {author} {\bibfnamefont {R.}~\bibnamefont {Cruz~Perez}}, \bibinfo {author} {\bibfnamefont {I.~O.}\ \bibnamefont {Yekinni}}, \bibinfo {author} {\bibfnamefont {B.}~\bibnamefont {Rothschild-Mancinelli}}, \bibinfo {author} {\bibfnamefont {G.~A.}\ \bibnamefont {Roberts~Kingman}}, \bibinfo {author} {\bibfnamefont {F.}~\bibnamefont {Wu}}, \bibinfo {author} {\bibfnamefont {J.}~\bibnamefont {Waeterschoot}}, \bibinfo {author} {\bibfnamefont {I.~A.}\ \bibnamefont {Ioannou}}, \bibinfo {author} {\bibfnamefont {M.~C.}\ \bibnamefont {Jewett}}, \bibinfo {author} {\bibfnamefont {A.~P.}\ \bibnamefont {Liu}}, \bibinfo {author} {\bibfnamefont {V.}~\bibnamefont
  {Noireaux}}, \bibinfo {author} {\bibfnamefont {C.}~\bibnamefont {Sorenson}}, \ and\ \bibinfo {author} {\bibfnamefont {K.~P.}\ \bibnamefont {Adamala}},\ }\href {\doibase 10.1021/acssynbio.3c00724} {\bibfield  {journal} {\bibinfo  {journal} {ACS Synthetic Biology}\ }\textbf {\bibinfo {volume} {13}},\ \bibinfo {pages} {974} (\bibinfo {year} {2024})}\BibitemShut {NoStop}%
\bibitem [{\citenamefont {Li}\ \emph {et~al.}(2025)\citenamefont {Li}, \citenamefont {Ilhamsyah}, \citenamefont {Tai},\ and\ \citenamefont {Shen}}]{https://doi.org/10.1002/adma.202414703}%
  \BibitemOpen
  \bibfield  {author} {\bibinfo {author} {\bibfnamefont {T.}~\bibnamefont {Li}}, \bibinfo {author} {\bibfnamefont {D.}~\bibnamefont {Ilhamsyah}}, \bibinfo {author} {\bibfnamefont {B.}~\bibnamefont {Tai}}, \ and\ \bibinfo {author} {\bibfnamefont {Y.}~\bibnamefont {Shen}},\ }\href {\doibase https://doi.org/10.1002/adma.202414703} {\bibfield  {journal} {\bibinfo  {journal} {Advanced Materials}\ }\textbf {\bibinfo {volume} {37}},\ \bibinfo {pages} {2414703} (\bibinfo {year} {2025})}\BibitemShut {NoStop}%
\bibitem [{\citenamefont {Yao}\ \emph {et~al.}(2024)\citenamefont {Yao}, \citenamefont {Feng},\ and\ \citenamefont {Yan}}]{YAO2024111170}%
  \BibitemOpen
  \bibfield  {author} {\bibinfo {author} {\bibfnamefont {T.}~\bibnamefont {Yao}}, \bibinfo {author} {\bibfnamefont {C.}~\bibnamefont {Feng}}, \ and\ \bibinfo {author} {\bibfnamefont {H.}~\bibnamefont {Yan}},\ }\href {\doibase https://doi.org/10.1016/j.microc.2024.111170} {\bibfield  {journal} {\bibinfo  {journal} {Microchemical Journal}\ }\textbf {\bibinfo {volume} {204}},\ \bibinfo {pages} {111170} (\bibinfo {year} {2024})}\BibitemShut {NoStop}%
\bibitem [{\citenamefont {Ruiz-Martínez}\ \emph {et~al.}(2025)\citenamefont {Ruiz-Martínez}, \citenamefont {Leermakers}, \citenamefont {Stoyanov},\ and\ \citenamefont {van~der Gucht}}]{doi:10.1021/acs.langmuir.5c00749}%
  \BibitemOpen
  \bibfield  {author} {\bibinfo {author} {\bibfnamefont {L.}~\bibnamefont {Ruiz-Martínez}}, \bibinfo {author} {\bibfnamefont {F.}~\bibnamefont {Leermakers}}, \bibinfo {author} {\bibfnamefont {S.}~\bibnamefont {Stoyanov}}, \ and\ \bibinfo {author} {\bibfnamefont {J.}~\bibnamefont {van~der Gucht}},\ }\href {\doibase 10.1021/acs.langmuir.5c00749} {\bibfield  {journal} {\bibinfo  {journal} {Langmuir}\ }\textbf {\bibinfo {volume} {41}},\ \bibinfo {pages} {11604} (\bibinfo {year} {2025})}\BibitemShut {NoStop}%
\bibitem [{\citenamefont {Scacchi}\ \emph {et~al.}(2025)\citenamefont {Scacchi}, \citenamefont {Rigoni}, \citenamefont {Haataja}, \citenamefont {Timonen},\ and\ \citenamefont {Sammalkorpi}}]{SCACCHI20251135}%
  \BibitemOpen
  \bibfield  {author} {\bibinfo {author} {\bibfnamefont {A.}~\bibnamefont {Scacchi}}, \bibinfo {author} {\bibfnamefont {C.}~\bibnamefont {Rigoni}}, \bibinfo {author} {\bibfnamefont {M.}~\bibnamefont {Haataja}}, \bibinfo {author} {\bibfnamefont {J.~V.}\ \bibnamefont {Timonen}}, \ and\ \bibinfo {author} {\bibfnamefont {M.}~\bibnamefont {Sammalkorpi}},\ }\href {\doibase https://doi.org/10.1016/j.jcis.2025.01.256} {\bibfield  {journal} {\bibinfo  {journal} {Journal of Colloid and Interface Science}\ }\textbf {\bibinfo {volume} {686}},\ \bibinfo {pages} {1135} (\bibinfo {year} {2025})}\BibitemShut {NoStop}%
\bibitem [{\citenamefont {Evans}(1979)}]{Evans01041979}%
  \BibitemOpen
  \bibfield  {author} {\bibinfo {author} {\bibfnamefont {R.}~\bibnamefont {Evans}},\ }\href {\doibase 10.1080/00018737900101365} {\bibfield  {journal} {\bibinfo  {journal} {Advances in Physics}\ }\textbf {\bibinfo {volume} {28}},\ \bibinfo {pages} {143} (\bibinfo {year} {1979})}\BibitemShut {NoStop}%
\bibitem [{\citenamefont {Evans}\ \emph {et~al.}(1993)\citenamefont {Evans}, \citenamefont {Henderson}, \citenamefont {Hoyle}, \citenamefont {Parry},\ and\ \citenamefont {Sabeur}}]{Evans01111993}%
  \BibitemOpen
  \bibfield  {author} {\bibinfo {author} {\bibfnamefont {R.}~\bibnamefont {Evans}}, \bibinfo {author} {\bibfnamefont {J.}~\bibnamefont {Henderson}}, \bibinfo {author} {\bibfnamefont {D.}~\bibnamefont {Hoyle}}, \bibinfo {author} {\bibfnamefont {A.}~\bibnamefont {Parry}}, \ and\ \bibinfo {author} {\bibfnamefont {Z.}~\bibnamefont {Sabeur}},\ }\href {\doibase 10.1080/00268979300102621} {\bibfield  {journal} {\bibinfo  {journal} {Molecular Physics}\ }\textbf {\bibinfo {volume} {80}},\ \bibinfo {pages} {755} (\bibinfo {year} {1993})}\BibitemShut {NoStop}%
\bibitem [{\citenamefont {Evans}(1992)}]{evans1992density}%
  \BibitemOpen
  \bibfield  {author} {\bibinfo {author} {\bibfnamefont {R.}~\bibnamefont {Evans}},\ }\href@noop {} {\bibfield  {journal} {\bibinfo  {journal} {Fundamentals of inhomogeneous fluids}\ }\textbf {\bibinfo {volume} {1}},\ \bibinfo {pages} {85} (\bibinfo {year} {1992})}\BibitemShut {NoStop}%
\bibitem [{\citenamefont {Evans}\ \emph {et~al.}(2016)\citenamefont {Evans}, \citenamefont {Oettel}, \citenamefont {Roth},\ and\ \citenamefont {Kahl}}]{Evans_2016}%
  \BibitemOpen
  \bibfield  {author} {\bibinfo {author} {\bibfnamefont {R.}~\bibnamefont {Evans}}, \bibinfo {author} {\bibfnamefont {M.}~\bibnamefont {Oettel}}, \bibinfo {author} {\bibfnamefont {R.}~\bibnamefont {Roth}}, \ and\ \bibinfo {author} {\bibfnamefont {G.}~\bibnamefont {Kahl}},\ }\href {\doibase 10.1088/0953-8984/28/24/240401} {\bibfield  {journal} {\bibinfo  {journal} {Journal of Physics: Condensed Matter}\ }\textbf {\bibinfo {volume} {28}},\ \bibinfo {pages} {240401} (\bibinfo {year} {2016})}\BibitemShut {NoStop}%
\bibitem [{\citenamefont {te~Vrugt}\ \emph {et~al.}(2020)\citenamefont {te~Vrugt}, \citenamefont {Löwen},\ and\ \citenamefont {Wittkowski}}]{Vrugt02042020}%
  \BibitemOpen
  \bibfield  {author} {\bibinfo {author} {\bibfnamefont {M.}~\bibnamefont {te~Vrugt}}, \bibinfo {author} {\bibfnamefont {H.}~\bibnamefont {Löwen}}, \ and\ \bibinfo {author} {\bibfnamefont {R.}~\bibnamefont {Wittkowski}},\ }\href {\doibase 10.1080/00018732.2020.1854965} {\bibfield  {journal} {\bibinfo  {journal} {Advances in Physics}\ }\textbf {\bibinfo {volume} {69}},\ \bibinfo {pages} {121} (\bibinfo {year} {2020})}\BibitemShut {NoStop}%
\bibitem [{\citenamefont {Tschopp}\ \emph {et~al.}(2020)\citenamefont {Tschopp}, \citenamefont {Vuijk}, \citenamefont {Sharma},\ and\ \citenamefont {Brader}}]{PhysRevE.102.042140}%
  \BibitemOpen
  \bibfield  {author} {\bibinfo {author} {\bibfnamefont {S.~M.}\ \bibnamefont {Tschopp}}, \bibinfo {author} {\bibfnamefont {H.~D.}\ \bibnamefont {Vuijk}}, \bibinfo {author} {\bibfnamefont {A.}~\bibnamefont {Sharma}}, \ and\ \bibinfo {author} {\bibfnamefont {J.~M.}\ \bibnamefont {Brader}},\ }\href {\doibase 10.1103/PhysRevE.102.042140} {\bibfield  {journal} {\bibinfo  {journal} {Physical Review E}\ }\textbf {\bibinfo {volume} {102}},\ \bibinfo {pages} {042140} (\bibinfo {year} {2020})}\BibitemShut {NoStop}%
\bibitem [{\citenamefont {Chen}\ \emph {et~al.}(2024)\citenamefont {Chen}, \citenamefont {Gao},\ and\ \citenamefont {Stinis}}]{10.1063/5.0175065}%
  \BibitemOpen
  \bibfield  {author} {\bibinfo {author} {\bibfnamefont {W.}~\bibnamefont {Chen}}, \bibinfo {author} {\bibfnamefont {P.}~\bibnamefont {Gao}}, \ and\ \bibinfo {author} {\bibfnamefont {P.}~\bibnamefont {Stinis}},\ }\href {\doibase 10.1063/5.0175065} {\bibfield  {journal} {\bibinfo  {journal} {Physics of Fluids}\ }\textbf {\bibinfo {volume} {36}},\ \bibinfo {pages} {017133} (\bibinfo {year} {2024})}\BibitemShut {NoStop}%
\bibitem [{\citenamefont {Tschopp}\ and\ \citenamefont {Brader}(2021)}]{PhysRevE.103.042103}%
  \BibitemOpen
  \bibfield  {author} {\bibinfo {author} {\bibfnamefont {S.~M.}\ \bibnamefont {Tschopp}}\ and\ \bibinfo {author} {\bibfnamefont {J.~M.}\ \bibnamefont {Brader}},\ }\href {\doibase 10.1103/PhysRevE.103.042103} {\bibfield  {journal} {\bibinfo  {journal} {Physical Review E}\ }\textbf {\bibinfo {volume} {103}},\ \bibinfo {pages} {042103} (\bibinfo {year} {2021})}\BibitemShut {NoStop}%
\bibitem [{\citenamefont {Sammüller}\ \emph {et~al.}(2024)\citenamefont {Sammüller}, \citenamefont {Hermann},\ and\ \citenamefont {Schmidt}}]{Sammüller_2024}%
  \BibitemOpen
  \bibfield  {author} {\bibinfo {author} {\bibfnamefont {F.}~\bibnamefont {Sammüller}}, \bibinfo {author} {\bibfnamefont {S.}~\bibnamefont {Hermann}}, \ and\ \bibinfo {author} {\bibfnamefont {M.}~\bibnamefont {Schmidt}},\ }\href {\doibase 10.1088/1361-648X/ad326f} {\bibfield  {journal} {\bibinfo  {journal} {Journal of Physics: Condensed Matter}\ }\textbf {\bibinfo {volume} {36}},\ \bibinfo {pages} {243002} (\bibinfo {year} {2024})}\BibitemShut {NoStop}%
\bibitem [{\citenamefont {Sammüller}\ \emph {et~al.}(2023)\citenamefont {Sammüller}, \citenamefont {Hermann}, \citenamefont {de~las Heras},\ and\ \citenamefont {Schmidt}}]{doi:10.1073/pnas.2312484120}%
  \BibitemOpen
  \bibfield  {author} {\bibinfo {author} {\bibfnamefont {F.}~\bibnamefont {Sammüller}}, \bibinfo {author} {\bibfnamefont {S.}~\bibnamefont {Hermann}}, \bibinfo {author} {\bibfnamefont {D.}~\bibnamefont {de~las Heras}}, \ and\ \bibinfo {author} {\bibfnamefont {M.}~\bibnamefont {Schmidt}},\ }\href {\doibase 10.1073/pnas.2312484120} {\bibfield  {journal} {\bibinfo  {journal} {Proceedings of the National Academy of Sciences}\ }\textbf {\bibinfo {volume} {120}},\ \bibinfo {pages} {e2312484120} (\bibinfo {year} {2023})}\BibitemShut {NoStop}%
\bibitem [{\citenamefont {Rosenfeld}(1989)}]{PhysRevLett.63.980}%
  \BibitemOpen
  \bibfield  {author} {\bibinfo {author} {\bibfnamefont {Y.}~\bibnamefont {Rosenfeld}},\ }\href {\doibase 10.1103/PhysRevLett.63.980} {\bibfield  {journal} {\bibinfo  {journal} {Physical Review Letters}\ }\textbf {\bibinfo {volume} {63}},\ \bibinfo {pages} {980} (\bibinfo {year} {1989})}\BibitemShut {NoStop}%
\bibitem [{\citenamefont {Oversteegen}\ and\ \citenamefont {Roth}(2005)}]{10.1063/1.1908765}%
  \BibitemOpen
  \bibfield  {author} {\bibinfo {author} {\bibfnamefont {S.~M.}\ \bibnamefont {Oversteegen}}\ and\ \bibinfo {author} {\bibfnamefont {R.}~\bibnamefont {Roth}},\ }\href {\doibase 10.1063/1.1908765} {\bibfield  {journal} {\bibinfo  {journal} {The Journal of Chemical Physics}\ }\textbf {\bibinfo {volume} {122}},\ \bibinfo {pages} {214502} (\bibinfo {year} {2005})}\BibitemShut {NoStop}%
\bibitem [{\citenamefont {Archer}\ \emph {et~al.}(2017)\citenamefont {Archer}, \citenamefont {Chacko},\ and\ \citenamefont {Evans}}]{10.1063/1.4993175}%
  \BibitemOpen
  \bibfield  {author} {\bibinfo {author} {\bibfnamefont {A.~J.}\ \bibnamefont {Archer}}, \bibinfo {author} {\bibfnamefont {B.}~\bibnamefont {Chacko}}, \ and\ \bibinfo {author} {\bibfnamefont {R.}~\bibnamefont {Evans}},\ }\href {\doibase 10.1063/1.4993175} {\bibfield  {journal} {\bibinfo  {journal} {The Journal of Chemical Physics}\ }\textbf {\bibinfo {volume} {147}},\ \bibinfo {pages} {034501} (\bibinfo {year} {2017})}\BibitemShut {NoStop}%
\bibitem [{\citenamefont {Schmidt}\ \emph {et~al.}(2000)\citenamefont {Schmidt}, \citenamefont {L\"owen}, \citenamefont {Brader},\ and\ \citenamefont {Evans}}]{PhysRevLett.85.1934}%
  \BibitemOpen
  \bibfield  {author} {\bibinfo {author} {\bibfnamefont {M.}~\bibnamefont {Schmidt}}, \bibinfo {author} {\bibfnamefont {H.}~\bibnamefont {L\"owen}}, \bibinfo {author} {\bibfnamefont {J.~M.}\ \bibnamefont {Brader}}, \ and\ \bibinfo {author} {\bibfnamefont {R.}~\bibnamefont {Evans}},\ }\href {\doibase 10.1103/PhysRevLett.85.1934} {\bibfield  {journal} {\bibinfo  {journal} {Physical Review Letters}\ }\textbf {\bibinfo {volume} {85}},\ \bibinfo {pages} {1934} (\bibinfo {year} {2000})}\BibitemShut {NoStop}%
\bibitem [{\citenamefont {Assis}\ \emph {et~al.}(2021)\citenamefont {Assis}, \citenamefont {Mageste}, \citenamefont {{de Lemos}}, \citenamefont {Orlando},\ and\ \citenamefont {Rodrigues}}]{ASSIS2021121697}%
  \BibitemOpen
  \bibfield  {author} {\bibinfo {author} {\bibfnamefont {R.~C.}\ \bibnamefont {Assis}}, \bibinfo {author} {\bibfnamefont {A.~B.}\ \bibnamefont {Mageste}}, \bibinfo {author} {\bibfnamefont {L.~R.}\ \bibnamefont {{de Lemos}}}, \bibinfo {author} {\bibfnamefont {R.~M.}\ \bibnamefont {Orlando}}, \ and\ \bibinfo {author} {\bibfnamefont {G.~D.}\ \bibnamefont {Rodrigues}},\ }\href {\doibase https://doi.org/10.1016/j.talanta.2020.121697} {\bibfield  {journal} {\bibinfo  {journal} {Talanta}\ }\textbf {\bibinfo {volume} {223}},\ \bibinfo {pages} {121697} (\bibinfo {year} {2021})}\BibitemShut {NoStop}%
\bibitem [{\citenamefont {Iqbal}\ \emph {et~al.}(2016)\citenamefont {Iqbal}, \citenamefont {Tao}, \citenamefont {Xie}, \citenamefont {Zhu}, \citenamefont {Chen}, \citenamefont {Wang}, \citenamefont {Huang}, \citenamefont {Peng}, \citenamefont {Sattar}, \citenamefont {Shabbir} \emph {et~al.}}]{iqbal2016aqueous}%
  \BibitemOpen
  \bibfield  {author} {\bibinfo {author} {\bibfnamefont {M.}~\bibnamefont {Iqbal}}, \bibinfo {author} {\bibfnamefont {Y.}~\bibnamefont {Tao}}, \bibinfo {author} {\bibfnamefont {S.}~\bibnamefont {Xie}}, \bibinfo {author} {\bibfnamefont {Y.}~\bibnamefont {Zhu}}, \bibinfo {author} {\bibfnamefont {D.}~\bibnamefont {Chen}}, \bibinfo {author} {\bibfnamefont {X.}~\bibnamefont {Wang}}, \bibinfo {author} {\bibfnamefont {L.}~\bibnamefont {Huang}}, \bibinfo {author} {\bibfnamefont {D.}~\bibnamefont {Peng}}, \bibinfo {author} {\bibfnamefont {A.}~\bibnamefont {Sattar}}, \bibinfo {author} {\bibfnamefont {M.~A.~B.}\ \bibnamefont {Shabbir}},  \emph {et~al.},\ }\href@noop {} {\bibfield  {journal} {\bibinfo  {journal} {Biological Procedures Online}\ }\textbf {\bibinfo {volume} {18}},\ \bibinfo {pages} {18} (\bibinfo {year} {2016})}\BibitemShut {NoStop}%
\bibitem [{\citenamefont {Rigoni}\ \emph {et~al.}(2022)\citenamefont {Rigoni}, \citenamefont {Beaune}, \citenamefont {Harnist}, \citenamefont {Sohrabi},\ and\ \citenamefont {Timonen}}]{rigoni2022ferrofluidic}%
  \BibitemOpen
  \bibfield  {author} {\bibinfo {author} {\bibfnamefont {C.}~\bibnamefont {Rigoni}}, \bibinfo {author} {\bibfnamefont {G.}~\bibnamefont {Beaune}}, \bibinfo {author} {\bibfnamefont {B.}~\bibnamefont {Harnist}}, \bibinfo {author} {\bibfnamefont {F.}~\bibnamefont {Sohrabi}}, \ and\ \bibinfo {author} {\bibfnamefont {J.~V.}\ \bibnamefont {Timonen}},\ }\href@noop {} {\bibfield  {journal} {\bibinfo  {journal} {Communications Materials}\ }\textbf {\bibinfo {volume} {3}},\ \bibinfo {pages} {26} (\bibinfo {year} {2022})}\BibitemShut {NoStop}%
\bibitem [{\citenamefont {Furuki}\ \emph {et~al.}(2024)\citenamefont {Furuki}, \citenamefont {Sakuta}, \citenamefont {Yanagisawa}, \citenamefont {Tabuchi}, \citenamefont {Kamo}, \citenamefont {Shimamoto},\ and\ \citenamefont {Yanagisawa}}]{furuki2024marangoni}%
  \BibitemOpen
  \bibfield  {author} {\bibinfo {author} {\bibfnamefont {T.}~\bibnamefont {Furuki}}, \bibinfo {author} {\bibfnamefont {H.}~\bibnamefont {Sakuta}}, \bibinfo {author} {\bibfnamefont {N.}~\bibnamefont {Yanagisawa}}, \bibinfo {author} {\bibfnamefont {S.}~\bibnamefont {Tabuchi}}, \bibinfo {author} {\bibfnamefont {A.}~\bibnamefont {Kamo}}, \bibinfo {author} {\bibfnamefont {D.~S.}\ \bibnamefont {Shimamoto}}, \ and\ \bibinfo {author} {\bibfnamefont {M.}~\bibnamefont {Yanagisawa}},\ }\href@noop {} {\bibfield  {journal} {\bibinfo  {journal} {ACS Applied Materials \& Interfaces}\ }\textbf {\bibinfo {volume} {16}},\ \bibinfo {pages} {43016} (\bibinfo {year} {2024})}\BibitemShut {NoStop}%
\bibitem [{\citenamefont {Holehouse}\ and\ \citenamefont {Alberti}(2025)}]{holehouse2025molecular}%
  \BibitemOpen
  \bibfield  {author} {\bibinfo {author} {\bibfnamefont {A.~S.}\ \bibnamefont {Holehouse}}\ and\ \bibinfo {author} {\bibfnamefont {S.}~\bibnamefont {Alberti}},\ }\href@noop {} {\bibfield  {journal} {\bibinfo  {journal} {Molecular cell}\ }\textbf {\bibinfo {volume} {85}},\ \bibinfo {pages} {290} (\bibinfo {year} {2025})}\BibitemShut {NoStop}%
\bibitem [{\citenamefont {Verlet}\ and\ \citenamefont {Weis}(1972)}]{PhysRevA.5.939}%
  \BibitemOpen
  \bibfield  {author} {\bibinfo {author} {\bibfnamefont {L.}~\bibnamefont {Verlet}}\ and\ \bibinfo {author} {\bibfnamefont {J.-J.}\ \bibnamefont {Weis}},\ }\href {\doibase 10.1103/PhysRevA.5.939} {\bibfield  {journal} {\bibinfo  {journal} {Physical Review A}\ }\textbf {\bibinfo {volume} {5}},\ \bibinfo {pages} {939} (\bibinfo {year} {1972})}\BibitemShut {NoStop}%
\bibitem [{\citenamefont {Toxvaerd}(1971)}]{10.1063/1.1676556}%
  \BibitemOpen
  \bibfield  {author} {\bibinfo {author} {\bibfnamefont {S.}~\bibnamefont {Toxvaerd}},\ }\href {\doibase 10.1063/1.1676556} {\bibfield  {journal} {\bibinfo  {journal} {The Journal of Chemical Physics}\ }\textbf {\bibinfo {volume} {55}},\ \bibinfo {pages} {3116} (\bibinfo {year} {1971})}\BibitemShut {NoStop}%
\bibitem [{\citenamefont {Henderson}(2021)}]{henderson2021fundamentals}%
  \BibitemOpen
  \bibfield  {author} {\bibinfo {author} {\bibfnamefont {D.}~\bibnamefont {Henderson}},\ }\href@noop {} {\emph {\bibinfo {title} {Fundamentals of inhomogeneous fluids}}}\ (\bibinfo  {publisher} {Crc Press},\ \bibinfo {year} {2021})\BibitemShut {NoStop}%
\bibitem [{\citenamefont {Weeks}\ \emph {et~al.}(1971)\citenamefont {Weeks}, \citenamefont {Chandler},\ and\ \citenamefont {Andersen}}]{10.1063/1.1674820}%
  \BibitemOpen
  \bibfield  {author} {\bibinfo {author} {\bibfnamefont {J.~D.}\ \bibnamefont {Weeks}}, \bibinfo {author} {\bibfnamefont {D.}~\bibnamefont {Chandler}}, \ and\ \bibinfo {author} {\bibfnamefont {H.~C.}\ \bibnamefont {Andersen}},\ }\href {\doibase 10.1063/1.1674820} {\bibfield  {journal} {\bibinfo  {journal} {The Journal of Chemical Physics}\ }\textbf {\bibinfo {volume} {54}},\ \bibinfo {pages} {5237} (\bibinfo {year} {1971})}\BibitemShut {NoStop}%
\bibitem [{\citenamefont {Louis}\ \emph {et~al.}(2000{\natexlab{a}})\citenamefont {Louis}, \citenamefont {Bolhuis}, \citenamefont {Hansen},\ and\ \citenamefont {Meijer}}]{Louis20002522}%
  \BibitemOpen
  \bibfield  {author} {\bibinfo {author} {\bibfnamefont {A.}~\bibnamefont {Louis}}, \bibinfo {author} {\bibfnamefont {P.}~\bibnamefont {Bolhuis}}, \bibinfo {author} {\bibfnamefont {J.}~\bibnamefont {Hansen}}, \ and\ \bibinfo {author} {\bibfnamefont {E.}~\bibnamefont {Meijer}},\ }\href {\doibase 10.1103/PhysRevLett.85.2522} {\bibfield  {journal} {\bibinfo  {journal} {Physical Review Letters}\ }\textbf {\bibinfo {volume} {85}},\ \bibinfo {pages} {2522 – 2525} (\bibinfo {year} {2000}{\natexlab{a}})}\BibitemShut {NoStop}%
\bibitem [{\citenamefont {Bolhuis}\ \emph {et~al.}(2001{\natexlab{a}})\citenamefont {Bolhuis}, \citenamefont {Louis}, \citenamefont {Hansen},\ and\ \citenamefont {Meijer}}]{Bolhuis20014296}%
  \BibitemOpen
  \bibfield  {author} {\bibinfo {author} {\bibfnamefont {P.}~\bibnamefont {Bolhuis}}, \bibinfo {author} {\bibfnamefont {A.}~\bibnamefont {Louis}}, \bibinfo {author} {\bibfnamefont {J.}~\bibnamefont {Hansen}}, \ and\ \bibinfo {author} {\bibfnamefont {E.}~\bibnamefont {Meijer}},\ }\href {\doibase 10.1063/1.1344606} {\bibfield  {journal} {\bibinfo  {journal} {The Journal of Chemical Physics}\ }\textbf {\bibinfo {volume} {114}},\ \bibinfo {pages} {4296 – 4311} (\bibinfo {year} {2001}{\natexlab{a}})}\BibitemShut {NoStop}%
\bibitem [{\citenamefont {Likos}(2001)}]{LIKOS2001267}%
  \BibitemOpen
  \bibfield  {author} {\bibinfo {author} {\bibfnamefont {C.~N.}\ \bibnamefont {Likos}},\ }\href {\doibase https://doi.org/10.1016/S0370-1573(00)00141-1} {\bibfield  {journal} {\bibinfo  {journal} {Physics Reports}\ }\textbf {\bibinfo {volume} {348}},\ \bibinfo {pages} {267} (\bibinfo {year} {2001})}\BibitemShut {NoStop}%
\bibitem [{\citenamefont {Götze}\ \emph {et~al.}(2004)\citenamefont {Götze}, \citenamefont {Harreis},\ and\ \citenamefont {Likos}}]{Götze20047761}%
  \BibitemOpen
  \bibfield  {author} {\bibinfo {author} {\bibfnamefont {I.}~\bibnamefont {Götze}}, \bibinfo {author} {\bibfnamefont {H.}~\bibnamefont {Harreis}}, \ and\ \bibinfo {author} {\bibfnamefont {C.}~\bibnamefont {Likos}},\ }\href {\doibase 10.1063/1.1689292} {\bibfield  {journal} {\bibinfo  {journal} {The Journal of Chemical Physics}\ }\textbf {\bibinfo {volume} {120}},\ \bibinfo {pages} {7761 – 7771} (\bibinfo {year} {2004})}\BibitemShut {NoStop}%
\bibitem [{\citenamefont {Götze}\ \emph {et~al.}(2006)\citenamefont {Götze}, \citenamefont {Archer},\ and\ \citenamefont {Likos}}]{10.1063/1.2172596}%
  \BibitemOpen
  \bibfield  {author} {\bibinfo {author} {\bibfnamefont {I.~O.}\ \bibnamefont {Götze}}, \bibinfo {author} {\bibfnamefont {A.~J.}\ \bibnamefont {Archer}}, \ and\ \bibinfo {author} {\bibfnamefont {C.~N.}\ \bibnamefont {Likos}},\ }\href {\doibase 10.1063/1.2172596} {\bibfield  {journal} {\bibinfo  {journal} {The Journal of Chemical Physics}\ }\textbf {\bibinfo {volume} {124}},\ \bibinfo {pages} {084901} (\bibinfo {year} {2006})}\BibitemShut {NoStop}%
\bibitem [{\citenamefont {Archer}\ and\ \citenamefont {Evans}(2001)}]{PhysRevE.64.041501}%
  \BibitemOpen
  \bibfield  {author} {\bibinfo {author} {\bibfnamefont {A.~J.}\ \bibnamefont {Archer}}\ and\ \bibinfo {author} {\bibfnamefont {R.}~\bibnamefont {Evans}},\ }\href {\doibase 10.1103/PhysRevE.64.041501} {\bibfield  {journal} {\bibinfo  {journal} {Physical Review E}\ }\textbf {\bibinfo {volume} {64}},\ \bibinfo {pages} {041501} (\bibinfo {year} {2001})}\BibitemShut {NoStop}%
\bibitem [{\citenamefont {Scacchi}\ \emph {et~al.}(2021{\natexlab{a}})\citenamefont {Scacchi}, \citenamefont {Nikkhah}, \citenamefont {Sammalkorpi},\ and\ \citenamefont {Ala-Nissila}}]{PhysRevResearch.3.L022008}%
  \BibitemOpen
  \bibfield  {author} {\bibinfo {author} {\bibfnamefont {A.}~\bibnamefont {Scacchi}}, \bibinfo {author} {\bibfnamefont {S.~J.}\ \bibnamefont {Nikkhah}}, \bibinfo {author} {\bibfnamefont {M.}~\bibnamefont {Sammalkorpi}}, \ and\ \bibinfo {author} {\bibfnamefont {T.}~\bibnamefont {Ala-Nissila}},\ }\href {\doibase 10.1103/PhysRevResearch.3.L022008} {\bibfield  {journal} {\bibinfo  {journal} {Physical Review Research}\ }\textbf {\bibinfo {volume} {3}},\ \bibinfo {pages} {L022008} (\bibinfo {year} {2021}{\natexlab{a}})}\BibitemShut {NoStop}%
\bibitem [{\citenamefont {Scacchi}\ \emph {et~al.}(2021{\natexlab{b}})\citenamefont {Scacchi}, \citenamefont {Sammalkorpi},\ and\ \citenamefont {Ala-Nissila}}]{10.1063/5.0053365}%
  \BibitemOpen
  \bibfield  {author} {\bibinfo {author} {\bibfnamefont {A.}~\bibnamefont {Scacchi}}, \bibinfo {author} {\bibfnamefont {M.}~\bibnamefont {Sammalkorpi}}, \ and\ \bibinfo {author} {\bibfnamefont {T.}~\bibnamefont {Ala-Nissila}},\ }\href {\doibase 10.1063/5.0053365} {\bibfield  {journal} {\bibinfo  {journal} {The Journal of Chemical Physics}\ }\textbf {\bibinfo {volume} {155}},\ \bibinfo {pages} {014904} (\bibinfo {year} {2021}{\natexlab{b}})}\BibitemShut {NoStop}%
\bibitem [{\citenamefont {Roth}(2010)}]{Roth_2010}%
  \BibitemOpen
  \bibfield  {author} {\bibinfo {author} {\bibfnamefont {R.}~\bibnamefont {Roth}},\ }\href {\doibase 10.1088/0953-8984/22/6/063102} {\bibfield  {journal} {\bibinfo  {journal} {Journal of Physics: Condensed Matter}\ }\textbf {\bibinfo {volume} {22}},\ \bibinfo {pages} {063102} (\bibinfo {year} {2010})}\BibitemShut {NoStop}%
\bibitem [{\citenamefont {Brader}\ \emph {et~al.}(2003)\citenamefont {Brader}, \citenamefont {Evans},\ and\ \citenamefont {Schmidt}}]{brader2003statistical}%
  \BibitemOpen
  \bibfield  {author} {\bibinfo {author} {\bibfnamefont {J.~M.}\ \bibnamefont {Brader}}, \bibinfo {author} {\bibfnamefont {R.}~\bibnamefont {Evans}}, \ and\ \bibinfo {author} {\bibfnamefont {M.}~\bibnamefont {Schmidt}},\ }\href@noop {} {\bibfield  {journal} {\bibinfo  {journal} {Molecular Physics}\ }\textbf {\bibinfo {volume} {101}},\ \bibinfo {pages} {3349} (\bibinfo {year} {2003})}\BibitemShut {NoStop}%
\bibitem [{\citenamefont {Schmidt}\ \emph {et~al.}(2003)\citenamefont {Schmidt}, \citenamefont {Denton},\ and\ \citenamefont {Brader}}]{schmidt2003fluid}%
  \BibitemOpen
  \bibfield  {author} {\bibinfo {author} {\bibfnamefont {M.}~\bibnamefont {Schmidt}}, \bibinfo {author} {\bibfnamefont {A.~R.}\ \bibnamefont {Denton}}, \ and\ \bibinfo {author} {\bibfnamefont {J.~M.}\ \bibnamefont {Brader}},\ }\href@noop {} {\bibfield  {journal} {\bibinfo  {journal} {The Journal of chemical physics}\ }\textbf {\bibinfo {volume} {118}},\ \bibinfo {pages} {1541} (\bibinfo {year} {2003})}\BibitemShut {NoStop}%
\bibitem [{\citenamefont {Reiss}(1992)}]{reiss1992statistical}%
  \BibitemOpen
  \bibfield  {author} {\bibinfo {author} {\bibfnamefont {H.}~\bibnamefont {Reiss}},\ }\href@noop {} {\bibfield  {journal} {\bibinfo  {journal} {The Journal of Physical Chemistry}\ }\textbf {\bibinfo {volume} {96}},\ \bibinfo {pages} {4736} (\bibinfo {year} {1992})}\BibitemShut {NoStop}%
\bibitem [{\citenamefont {Warren}\ \emph {et~al.}(1995)\citenamefont {Warren}, \citenamefont {Ilett},\ and\ \citenamefont {Poon}}]{warren1995effect}%
  \BibitemOpen
  \bibfield  {author} {\bibinfo {author} {\bibfnamefont {P.}~\bibnamefont {Warren}}, \bibinfo {author} {\bibfnamefont {S.}~\bibnamefont {Ilett}}, \ and\ \bibinfo {author} {\bibfnamefont {W.}~\bibnamefont {Poon}},\ }\href@noop {} {\bibfield  {journal} {\bibinfo  {journal} {Physical Review E}\ }\textbf {\bibinfo {volume} {52}},\ \bibinfo {pages} {5205} (\bibinfo {year} {1995})}\BibitemShut {NoStop}%
\bibitem [{\citenamefont {Melnyk}\ \emph {et~al.}(2022)\citenamefont {Melnyk}, \citenamefont {Trokhymchuk},\ and\ \citenamefont {Baumketner}}]{MELNYK2022120672}%
  \BibitemOpen
  \bibfield  {author} {\bibinfo {author} {\bibfnamefont {R.}~\bibnamefont {Melnyk}}, \bibinfo {author} {\bibfnamefont {A.}~\bibnamefont {Trokhymchuk}}, \ and\ \bibinfo {author} {\bibfnamefont {A.}~\bibnamefont {Baumketner}},\ }\href {\doibase https://doi.org/10.1016/j.molliq.2022.120672} {\bibfield  {journal} {\bibinfo  {journal} {Journal of Molecular Liquids}\ }\textbf {\bibinfo {volume} {368}},\ \bibinfo {pages} {120672} (\bibinfo {year} {2022})}\BibitemShut {NoStop}%
\bibitem [{\citenamefont {Lekkerkerker}\ \emph {et~al.}(1992)\citenamefont {Lekkerkerker}, \citenamefont {Poon}, \citenamefont {Pusey}, \citenamefont {Stroobants},\ and\ \citenamefont {Warren}}]{Lekkerkerker_1992}%
  \BibitemOpen
  \bibfield  {author} {\bibinfo {author} {\bibfnamefont {H.~N.~W.}\ \bibnamefont {Lekkerkerker}}, \bibinfo {author} {\bibfnamefont {W.~C.-K.}\ \bibnamefont {Poon}}, \bibinfo {author} {\bibfnamefont {P.~N.}\ \bibnamefont {Pusey}}, \bibinfo {author} {\bibfnamefont {A.}~\bibnamefont {Stroobants}}, \ and\ \bibinfo {author} {\bibfnamefont {P.~B.}\ \bibnamefont {Warren}},\ }\href {\doibase 10.1209/0295-5075/20/6/015} {\bibfield  {journal} {\bibinfo  {journal} {Europhysics Letters}\ }\textbf {\bibinfo {volume} {20}},\ \bibinfo {pages} {559} (\bibinfo {year} {1992})}\BibitemShut {NoStop}%
\bibitem [{\citenamefont {Bolhuis}\ \emph {et~al.}(2001{\natexlab{b}})\citenamefont {Bolhuis}, \citenamefont {Louis}, \citenamefont {Hansen},\ and\ \citenamefont {Meijer}}]{10.1063/1.1344606}%
  \BibitemOpen
  \bibfield  {author} {\bibinfo {author} {\bibfnamefont {P.~G.}\ \bibnamefont {Bolhuis}}, \bibinfo {author} {\bibfnamefont {A.~A.}\ \bibnamefont {Louis}}, \bibinfo {author} {\bibfnamefont {J.~P.}\ \bibnamefont {Hansen}}, \ and\ \bibinfo {author} {\bibfnamefont {E.~J.}\ \bibnamefont {Meijer}},\ }\href {\doibase 10.1063/1.1344606} {\bibfield  {journal} {\bibinfo  {journal} {The Journal of Chemical Physics}\ }\textbf {\bibinfo {volume} {114}},\ \bibinfo {pages} {4296} (\bibinfo {year} {2001}{\natexlab{b}})}\BibitemShut {NoStop}%
\bibitem [{\citenamefont {Louis}\ \emph {et~al.}(2000{\natexlab{b}})\citenamefont {Louis}, \citenamefont {Bolhuis},\ and\ \citenamefont {Hansen}}]{PhysRevE.62.7961}%
  \BibitemOpen
  \bibfield  {author} {\bibinfo {author} {\bibfnamefont {A.~A.}\ \bibnamefont {Louis}}, \bibinfo {author} {\bibfnamefont {P.~G.}\ \bibnamefont {Bolhuis}}, \ and\ \bibinfo {author} {\bibfnamefont {J.~P.}\ \bibnamefont {Hansen}},\ }\href {\doibase 10.1103/PhysRevE.62.7961} {\bibfield  {journal} {\bibinfo  {journal} {Physical Review E}\ }\textbf {\bibinfo {volume} {62}},\ \bibinfo {pages} {7961} (\bibinfo {year} {2000}{\natexlab{b}})}\BibitemShut {NoStop}%
\bibitem [{\citenamefont {Barker}\ and\ \citenamefont {Henderson}(1967)}]{10.1063/1.1701689}%
  \BibitemOpen
  \bibfield  {author} {\bibinfo {author} {\bibfnamefont {J.~A.}\ \bibnamefont {Barker}}\ and\ \bibinfo {author} {\bibfnamefont {D.}~\bibnamefont {Henderson}},\ }\href {\doibase 10.1063/1.1701689} {\bibfield  {journal} {\bibinfo  {journal} {The Journal of Chemical Physics}\ }\textbf {\bibinfo {volume} {47}},\ \bibinfo {pages} {4714} (\bibinfo {year} {1967})}\BibitemShut {NoStop}%
\bibitem [{\citenamefont {Farag}\ \emph {et~al.}(2023)\citenamefont {Farag}, \citenamefont {Borcherds}, \citenamefont {Bremer}, \citenamefont {Mittag},\ and\ \citenamefont {Pappu}}]{farag2023phase}%
  \BibitemOpen
  \bibfield  {author} {\bibinfo {author} {\bibfnamefont {M.}~\bibnamefont {Farag}}, \bibinfo {author} {\bibfnamefont {W.~M.}\ \bibnamefont {Borcherds}}, \bibinfo {author} {\bibfnamefont {A.}~\bibnamefont {Bremer}}, \bibinfo {author} {\bibfnamefont {T.}~\bibnamefont {Mittag}}, \ and\ \bibinfo {author} {\bibfnamefont {R.~V.}\ \bibnamefont {Pappu}},\ }\href@noop {} {\bibfield  {journal} {\bibinfo  {journal} {Nature Communications}\ }\textbf {\bibinfo {volume} {14}},\ \bibinfo {pages} {5527} (\bibinfo {year} {2023})}\BibitemShut {NoStop}%
\bibitem [{\citenamefont {Kamo}\ \emph {et~al.}(2025)\citenamefont {Kamo}, \citenamefont {Nikoubashman},\ and\ \citenamefont {Yanagisawa}}]{kamo2025impact}%
  \BibitemOpen
  \bibfield  {author} {\bibinfo {author} {\bibfnamefont {A.}~\bibnamefont {Kamo}}, \bibinfo {author} {\bibfnamefont {A.}~\bibnamefont {Nikoubashman}}, \ and\ \bibinfo {author} {\bibfnamefont {M.}~\bibnamefont {Yanagisawa}},\ }\href@noop {} {\bibfield  {journal} {\bibinfo  {journal} {The Journal of Physical Chemistry B}\ }\textbf {\bibinfo {volume} {129}},\ \bibinfo {pages} {3263} (\bibinfo {year} {2025})}\BibitemShut {NoStop}%
\bibitem [{\citenamefont {Velasco}\ and\ \citenamefont {Tarazona}(1989)}]{velasco1989wetting}%
  \BibitemOpen
  \bibfield  {author} {\bibinfo {author} {\bibfnamefont {E.}~\bibnamefont {Velasco}}\ and\ \bibinfo {author} {\bibfnamefont {P.}~\bibnamefont {Tarazona}},\ }\href@noop {} {\bibfield  {journal} {\bibinfo  {journal} {Journal of Chemical Physics}\ }\textbf {\bibinfo {volume} {91}},\ \bibinfo {pages} {7916} (\bibinfo {year} {1989})}\BibitemShut {NoStop}%
\bibitem [{\citenamefont {Bruno}\ \emph {et~al.}(1987)\citenamefont {Bruno}, \citenamefont {Caccamo},\ and\ \citenamefont {Tarazona}}]{bruno1987wetting}%
  \BibitemOpen
  \bibfield  {author} {\bibinfo {author} {\bibfnamefont {E.}~\bibnamefont {Bruno}}, \bibinfo {author} {\bibfnamefont {C.}~\bibnamefont {Caccamo}}, \ and\ \bibinfo {author} {\bibfnamefont {P.}~\bibnamefont {Tarazona}},\ }\href@noop {} {\bibfield  {journal} {\bibinfo  {journal} {Physical Review A}\ }\textbf {\bibinfo {volume} {35}},\ \bibinfo {pages} {1210} (\bibinfo {year} {1987})}\BibitemShut {NoStop}%
\bibitem [{\citenamefont {Edison}\ \emph {et~al.}(2015{\natexlab{a}})\citenamefont {Edison}, \citenamefont {Tasios}, \citenamefont {Belli}, \citenamefont {Evans}, \citenamefont {Van~Roij},\ and\ \citenamefont {Dijkstra}}]{edison2015critical}%
  \BibitemOpen
  \bibfield  {author} {\bibinfo {author} {\bibfnamefont {J.~R.}\ \bibnamefont {Edison}}, \bibinfo {author} {\bibfnamefont {N.}~\bibnamefont {Tasios}}, \bibinfo {author} {\bibfnamefont {S.}~\bibnamefont {Belli}}, \bibinfo {author} {\bibfnamefont {R.}~\bibnamefont {Evans}}, \bibinfo {author} {\bibfnamefont {R.}~\bibnamefont {Van~Roij}}, \ and\ \bibinfo {author} {\bibfnamefont {M.}~\bibnamefont {Dijkstra}},\ }\href@noop {} {\bibfield  {journal} {\bibinfo  {journal} {Physical Review Letters}\ }\textbf {\bibinfo {volume} {114}},\ \bibinfo {pages} {038301} (\bibinfo {year} {2015}{\natexlab{a}})}\BibitemShut {NoStop}%
\bibitem [{\citenamefont {Edison}\ \emph {et~al.}(2015{\natexlab{b}})\citenamefont {Edison}, \citenamefont {Belli}, \citenamefont {Evans}, \citenamefont {Van~Roij},\ and\ \citenamefont {Dijkstra}}]{edison2015phase}%
  \BibitemOpen
  \bibfield  {author} {\bibinfo {author} {\bibfnamefont {J.~R.}\ \bibnamefont {Edison}}, \bibinfo {author} {\bibfnamefont {S.}~\bibnamefont {Belli}}, \bibinfo {author} {\bibfnamefont {R.}~\bibnamefont {Evans}}, \bibinfo {author} {\bibfnamefont {R.}~\bibnamefont {Van~Roij}}, \ and\ \bibinfo {author} {\bibfnamefont {M.}~\bibnamefont {Dijkstra}},\ }\href@noop {} {\bibfield  {journal} {\bibinfo  {journal} {Molecular Physics}\ }\textbf {\bibinfo {volume} {113}},\ \bibinfo {pages} {2546} (\bibinfo {year} {2015}{\natexlab{b}})}\BibitemShut {NoStop}%
\bibitem [{\citenamefont {Archer}\ \emph {et~al.}(2024)\citenamefont {Archer}, \citenamefont {Goddard}, \citenamefont {Sibley}, \citenamefont {Rawlings}, \citenamefont {Broadhurst}, \citenamefont {Ouali},\ and\ \citenamefont {Fairhurst}}]{D4SM00332B}%
  \BibitemOpen
  \bibfield  {author} {\bibinfo {author} {\bibfnamefont {A.~J.}\ \bibnamefont {Archer}}, \bibinfo {author} {\bibfnamefont {B.~D.}\ \bibnamefont {Goddard}}, \bibinfo {author} {\bibfnamefont {D.~N.}\ \bibnamefont {Sibley}}, \bibinfo {author} {\bibfnamefont {J.~T.}\ \bibnamefont {Rawlings}}, \bibinfo {author} {\bibfnamefont {R.}~\bibnamefont {Broadhurst}}, \bibinfo {author} {\bibfnamefont {F.~F.}\ \bibnamefont {Ouali}}, \ and\ \bibinfo {author} {\bibfnamefont {D.~J.}\ \bibnamefont {Fairhurst}},\ }\href {\doibase 10.1039/D4SM00332B} {\bibfield  {journal} {\bibinfo  {journal} {Soft Matter}\ }\textbf {\bibinfo {volume} {20}},\ \bibinfo {pages} {5889} (\bibinfo {year} {2024})}\BibitemShut {NoStop}%
\bibitem [{\citenamefont {Ma}\ \emph {et~al.}(2015)\citenamefont {Ma}, \citenamefont {Acosta}, \citenamefont {Whitney}, \citenamefont {Podgornik}, \citenamefont {Steinmetz}, \citenamefont {French},\ and\ \citenamefont {Parsegian}}]{ma2015determination}%
  \BibitemOpen
  \bibfield  {author} {\bibinfo {author} {\bibfnamefont {Y.}~\bibnamefont {Ma}}, \bibinfo {author} {\bibfnamefont {D.~M.}\ \bibnamefont {Acosta}}, \bibinfo {author} {\bibfnamefont {J.~R.}\ \bibnamefont {Whitney}}, \bibinfo {author} {\bibfnamefont {R.}~\bibnamefont {Podgornik}}, \bibinfo {author} {\bibfnamefont {N.~F.}\ \bibnamefont {Steinmetz}}, \bibinfo {author} {\bibfnamefont {R.~H.}\ \bibnamefont {French}}, \ and\ \bibinfo {author} {\bibfnamefont {V.~A.}\ \bibnamefont {Parsegian}},\ }\href@noop {} {\bibfield  {journal} {\bibinfo  {journal} {Journal of biological physics}\ }\textbf {\bibinfo {volume} {41}},\ \bibinfo {pages} {85} (\bibinfo {year} {2015})}\BibitemShut {NoStop}%
\bibitem [{\citenamefont {Schmidt}(2004)}]{schmidt2004rosenfeld}%
  \BibitemOpen
  \bibfield  {author} {\bibinfo {author} {\bibfnamefont {M.}~\bibnamefont {Schmidt}},\ }\href@noop {} {\bibfield  {journal} {\bibinfo  {journal} {Journal of Physics: Condensed Matter}\ }\textbf {\bibinfo {volume} {16}},\ \bibinfo {pages} {L351} (\bibinfo {year} {2004})}\BibitemShut {NoStop}%
\bibitem [{\citenamefont {H{\"a}rtel}\ \emph {et~al.}(2015)\citenamefont {H{\"a}rtel}, \citenamefont {Janssen}, \citenamefont {Samin},\ and\ \citenamefont {Van~Roij}}]{hartel2015fundamental}%
  \BibitemOpen
  \bibfield  {author} {\bibinfo {author} {\bibfnamefont {A.}~\bibnamefont {H{\"a}rtel}}, \bibinfo {author} {\bibfnamefont {M.}~\bibnamefont {Janssen}}, \bibinfo {author} {\bibfnamefont {S.}~\bibnamefont {Samin}}, \ and\ \bibinfo {author} {\bibfnamefont {R.}~\bibnamefont {Van~Roij}},\ }\href@noop {} {\bibfield  {journal} {\bibinfo  {journal} {Journal of Physics: Condensed Matter}\ }\textbf {\bibinfo {volume} {27}},\ \bibinfo {pages} {194129} (\bibinfo {year} {2015})}\BibitemShut {NoStop}%
\bibitem [{\citenamefont {Wittmann}\ \emph {et~al.}(2016)\citenamefont {Wittmann}, \citenamefont {Marechal},\ and\ \citenamefont {Mecke}}]{wittmann2016fundamental}%
  \BibitemOpen
  \bibfield  {author} {\bibinfo {author} {\bibfnamefont {R.}~\bibnamefont {Wittmann}}, \bibinfo {author} {\bibfnamefont {M.}~\bibnamefont {Marechal}}, \ and\ \bibinfo {author} {\bibfnamefont {K.}~\bibnamefont {Mecke}},\ }\href@noop {} {\bibfield  {journal} {\bibinfo  {journal} {Journal of Physics: Condensed Matter}\ }\textbf {\bibinfo {volume} {28}},\ \bibinfo {pages} {244003} (\bibinfo {year} {2016})}\BibitemShut {NoStop}%
\end{thebibliography}%

\end{document}